\documentclass[]{IAC} 

\usepackage{gensymb}
\usepackage[english]{babel}
\usepackage[utf8]{inputenc}
\usepackage{newtxtext,newtxmath} 
\usepackage{enumitem} 
\let\savewidebar\widebar
\let\widebar\relax
\let\savebigtimes\bigtimes
\let\bigtimes\relax
\let\savedegree\degree
\let\degree\relax
\usepackage{mathabx}
\let\degree\savedegree
\let\bigtimes\savebigtimes
\let\widebar\savewidebar

\usepackage[dvipsnames]{xcolor}
\usepackage{tikz}
\usetikzlibrary{positioning, matrix, decorations.pathreplacing, calligraphy, shapes.geometric}
\usetikzlibrary{arrows,automata}
\usepackage{tikz-3dplot}
\usepackage[hidelinks]{hyperref}
\usepackage{cleveref}

\usepackage{algorithmic}
\usepackage{algorithm}
\usepackage{cite}
\usepackage{tabularray}
\usepackage{nicefrac}
\usepackage{siunitx}
\usepackage{fancyhdr}
\makeatletter
\newcommand\fs@nobottomruled{\def\@fs@cfont{\bfseries}\let\@fs@capt\floatc@ruled
  \def\@fs@pre{\hrule height.8pt depth0pt \kern2pt}%
  \def\@fs@post{}
  \def\@fs@mid{\kern2pt\hrule\kern2pt}%
  \let\@fs@iftopcapt\iftrue}
\makeatother

\floatstyle{nobottomruled}
\restylefloat{algorithm}

\makeatletter
\newcommand\fs@noruled{\def\@fs@cfont{\bfseries}\let\@fs@capt\floatc@ruled
  \def\@fs@post{}
  \let\@fs@iftopcapt\iftrue}
\makeatother

\addto\captionsenglish{%

}

\begin{document}


\title{Multiple-Arc Optimization of Low-Thrust Earth-Moon Orbit Transfers Leveraging Implicit Costate Transformation}
%

\IACauthor{Alessandro Beolchi}{Department of Aerospace Engineering, Khalifa University of Science and Technology, Abu Dhabi, P.O. Box 127788, United Arab Emirates; 100064448@ku.ac.ae\\ Faculty of Civil and Industrial Engineering, Sapienza Università di Roma, Rome, Italy; beolchi.1802082@studenti.uniroma1.it}
\IACauthor{Mauro Pontani}{Department of Astronautical, Electrical, and Energy Engineering, Sapienza Università di Roma, Rome, Italy; mauro.pontani@uniroma1.it}
\IACauthor{Chiara Pozzi}{Department of Aerospace Engineering, Khalifa University of Science and Technology, Abu Dhabi, P.O. Box 127788, United Arab Emirates; 100064456@ku.ac.ae\\ Faculty of Civil and Industrial Engineering, Sapienza Università di Roma, Rome, Italy; pozzi.1791751@studenti.uniroma1.it}
\IACauthor{Elena Fantino}{Department of Aerospace Engineering, Khalifa University of Science and Technology, Abu Dhabi, P.O. Box 127788, United Arab Emirates; elena.fantino@ku.ac.ae}

\abstract{This work focuses on minimum-time low-thrust orbit transfers from a prescribed low Earth orbit to a specified low lunar orbit. The well-established indirect formulation of minimum-time orbit transfers is extended to a multibody dynamical framework, with initial and final orbits around two distinct primaries. To do this, different representations, useful for describing orbit dynamics, are introduced, i.e., modified equinoctial elements (MEE) and Cartesian coordinates (CC). Use of two sets of MEE, relative to either Earth or Moon, allows simple writing of the boundary conditions about the two celestial bodies, but requires the formulation of a multiple-arc trajectory optimization problem, including two legs: (a) geocentric leg and (b) selenocentric leg. In the numerical solution process, the transition between the two MEE representations uses CC, which play the role of convenient intermediate, matching variables. The multiple-arc formulation at hand leads to identifying a set of intermediate necessary conditions for optimality, at the transition between the two legs. This research proves that a closed-form solution to these intermediate conditions exists, leveraging implicit costate transformation. As a result, the parameter set for an indirect algorithm retains the reduced size of the typical set associated with a single-arc optimization problem. The indirect heuristic technique, based on the joint use of the necessary conditions and a heuristic algorithm (i.e., differential evolution in this study) is proposed as the numerical solution method, together with the definition of a layered fitness function, aimed at facilitating convergence. The minimum-time trajectory of interest is sought in a high-fidelity dynamical framework, with the use of planetary ephemeris and the inclusion of the simultaneous gravitational action of Sun, Earth, and Moon, along the entire transfer path. The numerical results unequivocally prove that the approach developed in this research is effective for determining minimum-time low-thrust Earth-Moon orbit transfers.\\

}

\maketitle \thispagestyle{fancy}


\section{INTRODUCTION} 

\indent In recent years, low-thrust electric propulsion has attracted an increasing interest by the scientific community, and has already found application in a variety of mission scenarios, such as NASA's Deep Space 1 \cite{rayman2000results} and ESA's SMART-1 \cite{racca2002smart}. The latter is the only low-thrust mission to the Moon to date. Due to the high values of the specific impulse, low thrust enables substantial propellant savings, at the expense of increasing the time of flight. As a consequence, the use of low thrust as the primary means of spacecraft propulsion is limited to robotic spaceflight. Pioneering studies on low-thrust trajectories about a single attracting body are due to {Edelbaum \cite{edelbaum1962use}}, one of the first scientists to point out the benefit of using low thrust in space. Most recently, extensive research on the same subject was carried out by {Petropoulos \cite{petropoulos2003simple}}, {Betts \cite{betts2015optimal}}, and {Kechichian \cite{kechichian1998low}}, to name a few. The interest in lunar exploration has considerably improved over the last 20 years, mainly due to evidence of water ice deposits in the lunar polar regions \cite{duke2002lunar}, later ascertained by Chandrayaan-1 \cite{pieters2009character} and Lunar Reconnaissance Orbiter \cite{colaprete2010detection}. Artemis, the latest and most ambitious space program by NASA, is entirely devoted to enhancing human presence and exploration endeavor on the surface of the Moon. To this end, NASA intends to assemble an orbiting space station in the vicinity of the Moon, which will be essential for supporting manned facilities on the lunar surface and to facilitate human exploration toward deep space \cite{smith2020artemis}. Within this novel and challenging framework, it is crucial to envision and ultimately design an orbital roadmap connecting key orbits in the Earth-Moon system. Thus, innovative and advantageous orbit transfer strategies in cislunar space are of great practical interest at the present time. 

\indent A large number of works in the recent scientific literature is devoted to investigating optimal low-thrust trajectories in the framework of the two-body problem, with both direct and indirect optimization {methods \cite{betts2000very,pontani2014optimal,pontani2020optimal}}. On the other hand, limited research was dedicated to the design of optimal paths in multibody environments. Minimum-fuel low-thrust Earth-Moon trajectories from low Earth orbit (LEO) to low lunar orbit (LLO) were investigated in the approximate framework of the circular restricted three-body problem (CR3BP) by Pierson and Kluever, assuming a fixed thrust-coast-thrust sequence and using a "hybrid" direct/indirect method, in both {planar \cite{pierson1994three}} and three-{dimensional \cite{kluever1995optimal}} geometries. A few years later, Herman and {Conway \cite{herman1998optimal}} employed the method of collocation with nonlinear programming to address minimum-fuel low-thrust Earth-Moon orbit transfers from an arbitrary Earth orbit to an arbitrary lunar orbit with continuous thrust. In this context, an ephemeris model was employed, but the gravitational action of the Sun was neglected. Betts and {Erb \cite{betts2003optimal}} tackled once more the problem of low-thrust Earth-Moon orbit transfers for an application representative of ESA's SMART-1 trajectory. The transcription method was used in conjunction with modified equinoctial elements (MEE), to retrieve the minimum-fuel path from an initial Ariane 5 elliptic Earth parking orbit to a final elliptic polar lunar orbit with periapse above the lunar south pole. Most recently, Pérez-Palau and {Epenoy \cite{perez2018fuel}} developed an indirect optimal control approach for minimum-fuel trajectories from LEO to different lunar orbits in the Sun-Earth-Moon bicircular restricted four-body problem. Other effective strategies for the preliminary analysis of minimum-fuel Earth-Moon orbit transfers were proposed by Mingotti, Topputo, and {Bernelli-Zazzera \cite{mingotti2007combined,mingotti2009low}}, by Zhang, Topputo, Bernelli-Zazzera, and Zhao \cite{zhang2015low}, and by Ozimek and {Howell \cite{ozimek2010low}}, leveraging low-thrust in conjunction with invariant manifolds \cite{gomez2001invariant}, in the framework of the CR3BP. Concurrently, remarkable works investigating low-energy transfers in multibody dynamical frameworks were conducted by Lantoine, Russell, and Campagnola \cite{lantoine2011optimization} and by Epenoy and Pérez-Palau \cite{epenoy2019lyapunov}. However, despite the prosperity of this recent research topic, a rigorous indirect optimization method for addressing multibody transfers with initial and final orbits around distinct primaries was not proposed yet. 
\indent The research that follows addresses the problem of identifying minimum-time low-thrust Earth-Moon orbit transfers, modeled in a multibody dynamical framework. Besides Earth and Moon, the Sun greatly affects the trajectories traveled by spacecraft in cislunar space. Therefore, the gravitational action of the Sun is included in this analysis. This study has the following objectives: (i) introduce a multiple-arc formulation for the trajectory optimization problem under consideration, using different representations for the spacecraft dynamical state, (ii) derive the extended set of necessary conditions for optimality, (iii) prove that all the multipoint conditions can be solved sequentially, also taking advantage of implicit costate transformation, (iv) propose an {advanced indirect heuristic algorithm \cite{pontani2015minimum}} as the numerical solution method, and (v) test the technique at hand in the challenging mission scenario of Earth-Moon orbit transfers in a high-fidelity dynamical framework, with the use of planetary ephemeris and the inclusion of the simultaneous gravitational action of Sun, Earth, and Moon, along the entire transfer path. 

\indent This work is organized as follows. In Section II, the reference frames and the two representations for the spacecraft trajectory (i.e., Cartesian Coordinates (CC) and MEE) are introduced, together with the governing equations for MEE. Details on the different mission legs and the related (coexisting) state representations are provided in Section III. The multiple-arc formulation of the problem is the main subject of Section IV, which also emphasizes the crucial role of the implicit costate transformation to solve (in closed form) the intermediate multipoint necessary conditions for optimality. Section V describes the solution strategy, which features an indirect heuristic method. Finally, Section VI is focused on the numerical solution of a variety of orbit transfers between prescribed orbits, an initial LEO and distinct final low-altitude lunar orbits.
\section{ORBIT DYNAMICS} 

\indent The topic of this research is three-dimensional spacecraft trajectory optimization using low-thrust propulsion. The space vehicle is modeled as a point mass and its orbital motion around the main attracting body (either Earth or Moon) is investigated under the following assumptions:
\begin{enumerate}[label=(\alph*)]
    \item\label{a} Sun, Earth, and Moon have spherical mass distribution; 
    \item\label{b} low thrust is steerable and throttleable and is provided by a constant-specific-impulse and thrust-limited engine; 
    \item\label{c} the positions of Sun, Earth, and Moon are retrieved from a high-fidelity ephemeris {model \cite{acton1996ancillary}}. 
\end{enumerate}

\noindent Assumption \ref{a} implies that the gravitational attraction obeys the inverse square law. This approximation is justified by the fact that, for most of the transfer, the spacecraft is relatively far from both Earth and Moon and, as a result, the effect of higher-order harmonics is negligible.



\indent Denoting $T$, $c$, $m$, and $m_{0}$ the thrust magnitude, the exhaust velocity of the propulsion system, the spacecraft mass, and its initial value, respectively, assumption \ref{b} yields the governing equation for the spacecraft mass ratio $m_R = \nicefrac{m}{m_{0}}$ 
\begin{equation}\label{mRdot}
    \dot{m}_R = \frac{\dot{m}}{m_{0}} = -\frac{T}{m_{0} \, c} = -\frac{u_{T}}{c}
\end{equation}
\noindent where $u_T = \nicefrac{T}{m_{0}}$. Thus, the thrust acceleration magnitude is $a_T = \nicefrac{u_T}{m_R}$. 

\indent The remainder of this section describes two representations for the spacecraft position and velocity, i.e., (1) Cartesian coordinates and (2) modified equinoctial elements. As a preliminary step, some useful reference frames are introduced. 

\subsection{Reference frames}

\indent The Earth-centered inertial reference frame (ECI J2000) and the Moon-centered inertial reference frame (MCI) are defined with respect to the heliocentric inertial reference frame (HCI), associated with vectrix
\begin{equation}
    \underline{\underline{\rm N}}_S = \begin{bmatrix} {\hat{c}_1}^S & {\hat{c}_2}^S & {\hat{c}_3}^S \end{bmatrix}
\end{equation}

\noindent where $\hat{c}_1^S$ is the vernal axis (corresponding to the intersection of the ecliptic plane with Earth's equatorial plane), $\hat{c}_3^S$ is aligned with Earth's orbital angular momentum $\hat{h}_{\Earth}$, and the triad ($\hat{c}_1^S$, $\hat{c}_2^S$, $\hat{c}_3^S$) is a right-handed sequence of unit vectors. 

\indent The ECI reference frame is associated with vectrix
\begin{equation}\label{ECI}
    \underline{\underline{\rm N}}_E = \begin{bmatrix} {\hat{c}_1}^{E} & {\hat{c}_2}^{E} & {\hat{c}_3}^{E} \end{bmatrix}
\end{equation}
\noindent where $\hat{c}_1^E$ and $\hat{c}_2^E$ lie on the Earth mean equatorial plane and are coplanar with $\hat{c}_1^S$, $\hat{c}_3^E$ points toward the Earth rotation axis and the triad ($\hat{c}_1^E$, $\hat{c}_2^E$, $\hat{c}_3^E$) is a right-handed sequence of unit vectors.

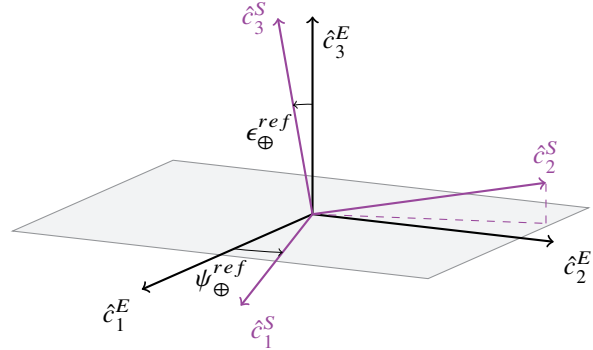
\begin{figure}[h]
    \centering
    \tdplotsetmaincoords{77}{117}
\begin{tikzpicture}[tdplot_main_coords]

\filldraw[
    draw=Gray,%
    fill=Gray!10,%
    ]          (-2,-3.1,0)
            -- (2.7,-3.1,0)
            -- (2.7,3.1,0)
            -- (-2,3.1,0)
            -- cycle;

\draw[thick,->] (0,0,0) -- (0,0,2.7) node[anchor=north west]{$\hat{c}_3^{E}$};
\draw[thick,->] (0,0,0) -- (0,3.6,0) node[anchor=north west]{$\hat{c}_2^{E}$};
\draw[thick,->] (0,0,0) -- (5,0,0) node[anchor=north east]{$\hat{c}_1^{E}$};
\draw[thick,->,Purple] (0,0,0) -- ({5.5*cos(17)},{5.5*sin(17)},0) node[anchor=north west]{$\hat{c}_1^{S}$};
\draw[thick,->,Purple] (0,0,0) -- ({2.7*sin(10)*sin(17)},{-2.7*sin(10)*cos(17)},{2.7*cos(10)}) node[anchor= east]{$\hat{c}_3^{S}$};
\draw[thick,->,Purple] (0,0,0) -- ({-3.2*cos(10)*sin(17)},{3.2*cos(10)*cos(17)},{3.2*sin(10)}) node[anchor= south]{$\hat{c}_2^{S}$};
\draw[dashed,Purple] (0,0,0) -- ({-3.2*cos(10)*sin(17)},{3.2*cos(10)*cos(17)},0);
\draw[dashed,Purple] ({-3.2*cos(10)*sin(17)},{3.2*cos(10)*cos(17)},0) -- ({-3.2*cos(10)*sin(17)},{3.2*cos(10)*cos(17)},{3.2*sin(10)});

\tdplotdrawarc[very thin,->]{(0,0,0)}{2.3}{0}{17}{anchor=north east}{$\psi_{\Earth}^{ref}$}
\tdplotdefinepoints(0,0,0)({5.5*cos(17)},{5.5*sin(17)},0)({-3.4*cos(10)*sin(17)},{3.4*cos(10)*cos(17)},{3.4*sin(10)})
\tdplotdefinepoints(0,0,0)(0,0,2.7)({2.7*sin(10)*sin(17)},{-2.7*sin(10)*cos(17)},{2.7*cos(10)})
\tdplotdrawpolytopearc[very thin,->]{1.5}{anchor= north east}{$\epsilon_{\Earth}^{ref}$}

\end{tikzpicture}
\caption{Reference frames HCI and ECI and related angles $\psi_{\Earth}^{ref}$ and $\epsilon_{\Earth}^{ref}$}
\label{Fig.HCIECI}
\end{figure}

\indent The ECI and the HCI reference frames are related through the angle $\psi_{\Earth}$ (separating ${\hat{c}_1}^S$ from ${\hat{c}_1}^E$) and the ecliptic obliquity angle $\epsilon_{\Earth}$ (separating ${\hat{c}_3}^S$ from ${\hat{c}_3}^E$), both taken at a reference epoch $t_{ref}$,
\begin{equation}\label{ECIHCI}
    \begin{bmatrix}
        {\hat{c}_1}^{S} \\ {\hat{c}_2}^{S} \\ {\hat{c}_3}^{S}
    \end{bmatrix}
    =
    {\rm \textbf{R}}_1\left(\epsilon_{\Earth}^{ref}\right) {\rm \textbf{R}}_3\left(\psi_{\Earth}^{ref}\right)
    \begin{bmatrix}
        {\hat{c}_1}^E \\ {\hat{c}_2}^E \\ {\hat{c}_3}^E
    \end{bmatrix}
\end{equation}
\noindent where ${\rm \textbf{R}}_j\left( \theta \right)$ denotes an elementary counterclockwise rotation about axis $j$ by a generic angle $\theta$. Figure \ref{Fig.HCIECI} depicts the two reference frames and the related angles.

\indent It is important to underline that ${\hat{c}_1}^S$ and ${\hat{c}_1}^E$ are exactly superimposed only at epoch J2000, that is the epoch at which the ECI reference frame is set. At any other epoch, the ecliptic plane does not intersect the ECI J2000 equatorial plane along ${\hat{c}_1}^E$ and, as a consequence, the angular displacement $\psi_{\Earth}$ arises. 

\begin{figure}[h]
    \centering
    \tdplotsetmaincoords{77}{117}
\begin{tikzpicture}[tdplot_main_coords]

\draw[thick,->] (0,0,0) -- (0,0,2.7) node[anchor=south]{$\hat{c}_3^{S}$};
\draw[thick,->,Orange] (0,0,0) -- ({2.7*sin(35)*sin(17)},{-2.7*sin(35)*cos(17)},{2.7*cos(35)}) node[anchor= east]{$\hat{h}_{\Moon}$};
\draw[thick,->,Blue] (0,0,0) -- ({-2.7*sin(20)*sin(17)},{2.7*sin(20)*cos(17)},{2.7*cos(20)}) node[anchor= west]{$\hat{c}_{3}^M$};

\tdplotdefinepoints(0,0,0)(0,0,2.7)({2.7*sin(35)*sin(17)},{-2.7*sin(35)*cos(17)},{2.7*cos(35)})
\tdplotdrawpolytopearc[very thin,->]{1.5}{anchor= north}{$\delta_{\Moon}$}
\tdplotdefinepoints(0,0,0)(0,0,2.7)({-2.7*sin(20)*sin(17)},{2.7*sin(20)*cos(17)},{2.7*cos(20)})
\tdplotdrawpolytopearc[very thin,->]{2}{anchor= north}{$\epsilon_{\Moon}$}
\tdplotdrawarc[thick,->,dotted,Orange]{(0,0,{2.7*cos(35)})}{{2.7*sin(35)}}{360}{0}{anchor= north}{}
\tdplotdrawarc[thick,->,dotted,Blue]{(0,0,{2.7*cos(20)})}{{2.7*sin(20)}}{360}{0}{anchor= north}{}

\end{tikzpicture}
\caption{Geometry of unit vectors $\hat{c}_{3}^S$, $\hat{h}_{\Moon}$, and $\hat{c}_{3}^M$ and related angles $\delta_{\Moon}$ and $\epsilon_{\Moon}$}
\label{Fig.Cones}
\end{figure}
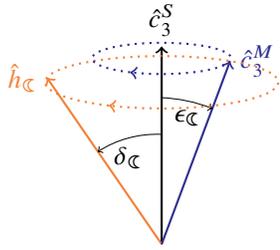

\indent According to Cassini's {laws\cite{peale1969generalized}}, the Moon rotation axis $\hat{z}_{\Moon}$, the Moon orbit angular momentum $\underrightarrow{\boldsymbol{h}}_{\Moon}$, and the normal to the ecliptic plane $\hat{c}_3^S$ all lie in the same plane. Moreover, the vectors $\hat{z}_{\Moon}$ and $\underrightarrow{\boldsymbol{h}}_{\Moon}$ are located at opposite sides of the ecliptic pole $\hat{c}_3^S$ and are both subject to clockwise precession, with a period of $18.6$ years, because of the Sun gravitational perturbation. Axis ${\hat{c}_3}^{M}$ is identified as the rotation axis $\hat{z}_{\Moon}$ at the reference epoch, while the remaining two axes ${\hat{c}_1}^{M}$ and ${\hat{c}_2}^{M}$ lie in the Moon equatorial plane, orthogonal to $\hat{z}_{\Moon}$. Figure \ref{Fig.Cones} illustrates the relative orientation of unit vectors ${\hat{c}_3}^{M}$, $\hat{c}_3^S$, and $\hat{h}_{\Moon}$. By definition, the axis ${\hat{c}_1}^{M}$ is chosen to be coplanar with the line that connects Earth and Moon at the reference epoch, lies on the plane orthogonal to ${\hat{c}_3}^{M}$, and is directed toward the far side of the Moon. Lastly, ${\hat{c}_2}^{M}$ is chosen such that (${\hat{c}_1}^{M}$, ${\hat{c}_2}^{M}$, ${\hat{c}_3}^{M}$) is a right-handed sequence of unit vectors.

\begin{figure}[h]
    \centering
    \tdplotsetmaincoords{77}{117}
\begin{tikzpicture}[tdplot_main_coords]

\filldraw[
    draw=Gray,%
    fill=Gray!10,%
    ]          (-2,-3.1,0)
            -- (2.7,-3.1,0)
            -- (2.7,3.1,0)
            -- (-2,3.1,0)
            -- cycle;

\draw[thick,->] (0,0,0) -- (0,0,2.7) node[anchor=north west]{$\hat{c}_3^{S}$};
\draw[thick,->] (0,0,0) -- (0,3.6,0) node[anchor=north west]{$\hat{c}_2^{S}$};
\draw[thick,->] (0,0,0) -- (5,0,0) node[anchor=north east]{$\hat{c}_1^{S}$};
\draw[thick,->,Orange] (0,0,0) -- ({5.5*cos(17)},{5.5*sin(17)},0) node[anchor=north west]{$\hat{N}_{\Moon}$};
\draw[thick,->,Orange] (0,0,0) -- ({2.7*sin(10)*sin(17)},{-2.7*sin(10)*cos(17)},{2.7*cos(10)}) node[anchor= east]{$\hat{h}_{\Moon}$};
\draw[thick,->,Orange] (0,0,0) -- ({-3.2*cos(10)*sin(17)},{3.2*cos(10)*cos(17)},{3.2*sin(10)}) node[anchor= south]{$\hat{M}_{\Moon}$};
\draw[dashed,Orange] (0,0,0) -- ({-3.2*cos(10)*sin(17)},{3.2*cos(10)*cos(17)},0);
\draw[dashed,Orange] ({-3.2*cos(10)*sin(17)},{3.2*cos(10)*cos(17)},0) -- ({-3.2*cos(10)*sin(17)},{3.2*cos(10)*cos(17)},{3.2*sin(10)});

\tdplotdrawarc[very thin,->]{(0,0,0)}{2.3}{0}{17}{anchor=north east}{$\psi_{\Moon}$}
\tdplotdefinepoints(0,0,0)({5.5*cos(17)},{5.5*sin(17)},0)({-3.4*cos(10)*sin(17)},{3.4*cos(10)*cos(17)},{3.4*sin(10)})
\tdplotdefinepoints(0,0,0)(0,0,2.7)({2.7*sin(10)*sin(17)},{-2.7*sin(10)*cos(17)},{2.7*cos(10)})
\tdplotdrawpolytopearc[very thin,->]{1.5}{anchor= north east}{$\delta_{\Moon}$}

\end{tikzpicture}
\caption{Reference frames MO and HCI and related angles $\psi_{\Moon}$ and $\epsilon_{\Moon}$}
\label{Fig.MOHCI}
\end{figure}
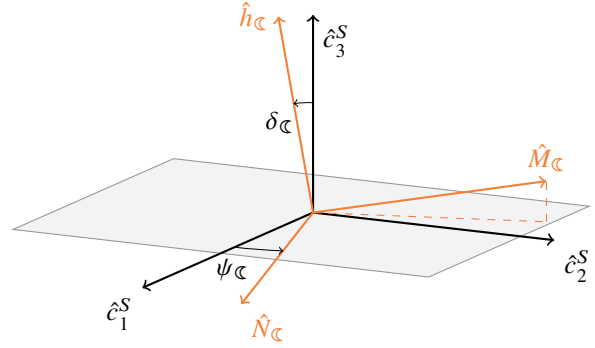

\indent As an intermediate step for the definition of the MCI reference frame, it is convenient to introduce the Moon orbital reference frame (MO)
\begin{equation}\label{MO}
    \underline{\underline{\rm R}}_{MO} = \begin{bmatrix} \hat{N}_{\Moon} & \hat{M}_{\Moon} & \hat{h}_{\Moon} \end{bmatrix}.
\end{equation}
\noindent The MO reference frame is related to the HCI reference frame through the Moon precession angle $\psi_{\Moon}$ (separating $\hat{N}_{\Moon}$ from ${\hat{c}_1}^S$) and the Moon orbital plane obliquity angle $\delta_{\Moon}$ (separating $\hat{h}_{\Moon}$ from ${\hat{c}_3}^S$)

\begin{equation}\label{MOHCI}
    \begin{bmatrix}
        \hat{N}_{\Moon} \\ \hat{M}_{\Moon} \\ \hat{h}_{\Moon}
    \end{bmatrix}
    =
    {\rm \textbf{R}}_1\left(\delta_{\Moon}\right) {\rm \textbf{R}}_3\left(\psi_{\Moon}\right)
    \begin{bmatrix}
        {\hat{c}_1}^S \\ {\hat{c}_2}^S \\ {\hat{c}_3}^S
    \end{bmatrix}
\end{equation}

\noindent where $\hat{N}_{\Moon}$ is the unit vector pointing toward the ascending node of the Moon orbit, $\hat{h}_{\Moon}$ is aligned with the Moon orbit angular momentum, and $\hat{M}_{\Moon}$ is chosen such that ($\hat{N}_{\Moon}$, $\hat{M}_{\Moon}$, $\hat{h}_{\Moon}$) is a right-handed sequence of unit vectors. Figure \ref{Fig.MOHCI} depicts the two reference frames and the related angles.

\indent The Moon-centered inertial reference frame (MCI) is closely related to the MO reference frame because $\hat{N}_{\Moon}$ is orthogonal to the plane containing $\hat{z}_{\Moon}$, $\underrightarrow{\boldsymbol{h}}_{\Moon}$, and $\hat{c}_3^S$. As a preliminary step, an auxiliary inertial reference frame $\underline{\underline{\rm N}}_{A} = \begin{bmatrix} {\hat{c}_1}^{A} & {\hat{c}_2}^{A} & {\hat{c}_3}^{A} \end{bmatrix}$ at the reference epoch $t_{ref}$ can be defined in relation to the HCI reference frame

\begin{equation}\label{AIHCI}
    \begin{bmatrix}
        {\hat{c}_1}^{A} \\ {\hat{c}_2}^{A} \\ {\hat{c}_3}^{A}
    \end{bmatrix}
    =
    {\rm \textbf{R}}_1\left(-\epsilon_{\Moon}\right) {\rm \textbf{R}}_3\left(\psi_{\Moon}^{ref}\right)
    \begin{bmatrix}
        {\hat{c}_1}^S \\ {\hat{c}_2}^S \\ {\hat{c}_3}^S
    \end{bmatrix}.
\end{equation}

\indent The Moon-centered inertial reference frame is obtained from the auxiliary inertial reference frame with a single elementary rotation about axis ${\hat{c}_3}^{A}$ by angle $\theta_{\Moon}^{ref}$, where $\theta_{\Moon}^{ref}$ is the angle that separates $\hat{N}_{\Moon}$ from the projection of the unit vector from the Earth to the Moon $\hat{r}_{\Earth \Moon}$ on the plane orthogonal to ${\hat{c}_3}^{A}$ ($= {\hat{c}_3}^{M}$), at the reference epoch

\begin{equation}\label{MCIAI}
    \begin{bmatrix}
        {\hat{c}_1}^{M} \\ {\hat{c}_2}^{M} \\ {\hat{c}_3}^{M}
    \end{bmatrix}
    =
    {\rm \textbf{R}}_3\left(\theta_{\Moon}^{ref}\right)
    \begin{bmatrix}
        {\hat{c}_1}^A \\ {\hat{c}_2}^A \\ {\hat{c}_3}^A
    \end{bmatrix}.
\end{equation}

\noindent Thus, it is possible to link the MCI reference frame to the HCI reference frame with a 3-1-3 sequence of elementary rotations by combining expressions (\ref{AIHCI}) and (\ref{MCIAI})
\begin{equation}\label{MCIHCI}
    \begin{bmatrix}
        {\hat{c}_1}^{M} \\ {\hat{c}_2}^{M} \\ {\hat{c}_3}^{M}
    \end{bmatrix}
    =
    {\rm \textbf{R}}_3\left(\theta_{\Moon}^{ref}\right) {\rm \textbf{R}}_1\left(-\epsilon_{\Moon}\right) {\rm \textbf{R}}_3\left(\psi_{\Moon}^{ref}\right) \begin{bmatrix}
        {\hat{c}_1}^S \\ {\hat{c}_2}^S \\ {\hat{c}_3}^S
    \end{bmatrix}.
\end{equation}
\indent Then, the MCI reference frame can be related to the ECI reference frame with a 3-1-3-1-3 sequence of elementary rotations, according to the combination of Eqs. (\ref{ECIHCI}) and (\ref{MCIHCI})
\begingroup
\thinmuskip=0mu
\medmuskip=0mu
\thickmuskip=0mu
\begin{equation}\label{MCIECI}
    \resizebox{.88\hsize}{!}{$\begin{bmatrix}
        {\hat{c}_1}^{M} \\ {\hat{c}_2}^{M} \\ {\hat{c}_3}^{M}
    \end{bmatrix}
    =
    {\rm \textbf{R}}_3\left(\theta_{\Moon}^{ref}\right) {\rm \textbf{R}}_1\left(-\epsilon_{\Moon}\right) {\rm \textbf{R}}_3\left(\psi_{\Moon}^{ref}\right) {\rm \textbf{R}}_1\left(\epsilon_{\Earth}^{ref}\right) {\rm \textbf{R}}_3\left(\psi_{\Earth}^{ref}\right) \begin{bmatrix}
        {\hat{c}_1}^E \\ {\hat{c}_2}^E \\ {\hat{c}_3}^E
    \end{bmatrix}.$}
\end{equation}
\endgroup
\noindent The corresponding rotation matrix binding the ECI reference frame to the MCI reference frame is 
\begingroup
\thinmuskip=0mu
\medmuskip=0mu
\thickmuskip=0mu
\begin{equation}\label{RMCIECI}
    \resizebox{.88\hsize}{!}{$\underset{\rm MCI \leftarrow ECI}{\rm \textbf{R}}
    =
    {\rm \textbf{R}}_3\left(\theta_{\Moon}^{ref}\right) {\rm \textbf{R}}_1\left(-\epsilon_{\Moon}\right) {\rm \textbf{R}}_3\left(\psi_{\Moon}^{ref}\right) {\rm \textbf{R}}_1\left(\epsilon_{\Earth}^{ref}\right) {\rm \textbf{R}}_3\left(\psi_{\Earth}^{ref}\right)$}
\end{equation}
\endgroup
\noindent where the angles $\psi_{\Earth}^{ref}$, $\epsilon_{\Earth}^{ref}$, $\psi_{\Moon}^{ref}$ and $\theta_{\Moon}^{ref}$ can all be retrieved from planetary ephemeris at the reference epoch. The remaining angle $\epsilon_{\Moon}$ is termed obliquity of the Moon equator and equals $1.62$ {degree\cite{peale1969generalized}}.
\begin{figure}[h]
    \centering
    \begin{tikzpicture}

    \node[rectangle, draw=Black, fill=Black!15, very thick, minimum size=1cm] at (0,0) (ECI) {$\underline{\underline{N}}_{E}$};

    \node[below = 1.1cm of ECI, circle, draw=Black, fill=Black, very thick, minimum size=1mm, scale=0.4] (ball1) {};

    \node[below = 1.1cm of ball1, rectangle, draw=Black, fill=Black!15, very thick, minimum size=1cm] (HCI) {$\underline{\underline{N}}_{S}$};

    \node[right = 1.7cm of HCI, circle, draw=Black, fill=Black, very thick, minimum size=1mm, scale=0.4] (ball2) {};

    \node[right = 3.4cm of HCI, rectangle, draw=Black, fill=Black!15, very thick, minimum size=1cm] (MO) {$\underline{\underline{R}}_{MO}$};

    \node[below = 1.1cm of HCI, circle, draw=Black, fill=Black, very thick, minimum size=1mm, scale=0.4] (ball3) {};

    \node[below = 2.2cm of HCI, rectangle, draw=Black, fill=Black!15, very thick, minimum size=1cm] (A) {$\underline{\underline{N}}_{A}$};

    \node[right = 3.4cm of A, rectangle, draw=Black, fill=Black!15, very thick, minimum size=1cm] (MCI) {$\underline{\underline{N}}_{M}$};
    
    \draw[->, draw=Black!90, very thick] (ECI) -- (ball1) node[midway, left] {${\rm \textbf{R}}_3\left(\psi_{\Earth}^{ref}\right)$};

    \draw[->, draw=Black!90, very thick] (ball1) -- (HCI) node[midway, left] {${\rm \textbf{R}}_1\left(\epsilon_{\Earth}^{ref}\right)$};

    \draw[->, draw=Black!90, very thick] (HCI) -- (ball2) node[midway, above] {${\rm \textbf{R}}_3\left(\psi_{\Moon}\right)$};

    \draw[->, draw=Black!90, very thick] (ball2) -- (MO) node[midway, above] {${\rm \textbf{R}}_1\left(\delta_{\Moon}\right)$};

    \draw[->, draw=Black!90, very thick] (HCI) -- (ball3) node[midway, left] {${\rm \textbf{R}}_3\left(\psi_{\Moon}^{ref}\right)$};

    \draw[->, draw=Black!90, very thick] (ball3) -- (A) node[midway, left] {${\rm \textbf{R}}_1\left(-\epsilon_{\Moon}\right)$};

    \draw[->, draw=Black!90, very thick] (A) -- (MCI) node[midway, above] {${\rm \textbf{R}}_3\left(\theta_{\Moon}^{ref}\right)$};
    
    \end{tikzpicture}
    \caption{Outline of the elementary rotations sequences linking the reference frames} 
    \label{Fig.RefFrameScheme}
\end{figure}
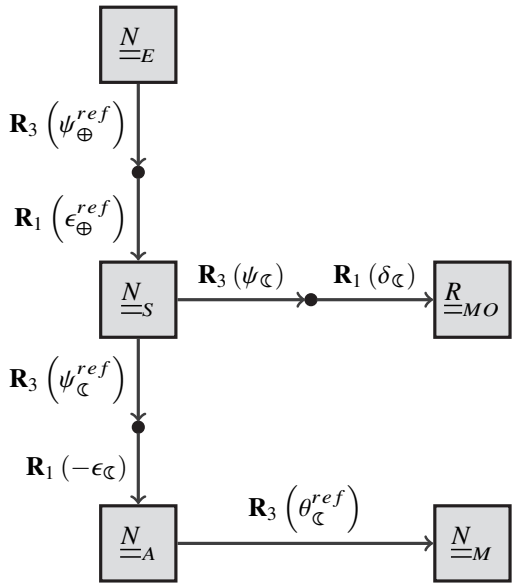

\indent In this work, the departure epoch is made to coincide with the reference epoch $t^{ref}$.

\indent Finally, the local vertical local horizontal frame represents a useful system that can only be defined in relation to a main attracting body $B$ and rotates together with the space vehicle. It is associated with vectrix 
\begin{equation}\label{LVLHB}
    \underline{\underline{\rm R}}_{LVLH_B} = \begin{bmatrix} {\hat{r}_B} & {\hat{\theta}_B} & {\hat{h}_B} \end{bmatrix}
\end{equation}
\noindent where $\hat{r}_B$ is aligned with the spacecraft position vector $\underrightarrow{\boldsymbol{r}}_B$ (taken from the center of mass of $B$), $\hat{h}_B$ points toward the spacecraft orbit angular momentum, whereas $\hat{\theta}_B$ is aligned with the projection of the satellite velocity $\underrightarrow{\boldsymbol{v}}$ into the local horizontal plane and it is such that ($\hat{r}_B$, $\hat{\theta}_B$, $\hat{h}_B$) is a right-handed sequence of unit vectors.

\indent Because either the Earth or the Moon can be regarded as the main attracting body, two distinct local vertical local horizontal reference frames $\underline{\underline{\rm R}}_{LVLH_E}$ and $\underline{\underline{\rm R}}_{LVLH_M}$ can be introduced.
\subsection{Cartesian coordinates}



\indent The spacecraft position and velocity can be identified in terms of Cartesian components along an inertial reference frame $\underline{\underline{N}}$. Variables $x$, $y$, and $z$ identify the spacecraft position in $\underline{\underline{N}}$, while the variables $v_x$, $v_y$, and $v_z$ represent the Cartesian coordinates of the velocity. The dynamical state of the space vehicle is identified by the state vector $\boldsymbol{y}$ defined as 
\begin{equation}
    \boldsymbol{y} = \begin{bmatrix}
        x & y & z & v_x & v_y & v_z & m_R
    \end{bmatrix}^T.
\end{equation} 
\subsection{Modified equinoctial elements}

\indent In several applications, modified equinoctial elements are preferred to classical orbit elements for numerical propagation, because they have the remarkable advantage of avoiding singularities when circular or equatorial orbits are encountered (or approached). Modified equinoctial elements can be introduced in terms of the classical orbit elements, i.e., semimajor axis $a$, eccentricity $e$, inclination $i$, right ascension of the ascending node (RAAN) $\Omega$, argument of periapsis $\omega$, and true anomaly $\theta_*$, as
\begin{equation}\label{COE2MEE}
    \begin{split}
        p &= a \, \left( 1 - e^2 \right) \;\;\;\;\;\;\;\;\; l = e \, \cos{(\Omega + \omega)}\\
        m &= e \, \sin{(\Omega + \omega)} \;\;\;\;\; n = \tan{\frac{i}{2}} \cos{\Omega}\\
        s &= \tan{\frac{i}{2}} \sin{\Omega} \;\;\;\;\;\;\;\;\; q = \Omega + \omega + \theta_{*}
    \end{split}
\end{equation}
\noindent where element $p$ is the semilatus rectum (\begingroup
\thinmuskip=0mu
\medmuskip=0mu
\thickmuskip=0mu$p > 0$\endgroup), elements $l$ and $m$ depend on eccentricity and longitude of the periapsis \begingroup
\thinmuskip=0mu
\medmuskip=0mu
\thickmuskip=0mu$\overline{\omega} = \Omega + \omega$\endgroup, elements $n$ and $s$ depend on inclination and RAAN, and element $q$ is the true longitude (\begingroup
\thinmuskip=0mu
\medmuskip=0mu
\thickmuskip=0mu$-\pi < q \leq \pi$\endgroup). These variables are nonsingular for all Keplerian trajectories, with the only exception of equatorial retrograde orbits (\begingroup
\thinmuskip=0mu
\medmuskip=0mu
\thickmuskip=0mu$i = \pi$\endgroup). Letting $x_6 {\equiv} q$ and $\boldsymbol{z} {=} \begin{bmatrix}
        x_1 & x_2 & x_3 & x_4 & x_5
    \end{bmatrix}^T {\equiv}\\ {\equiv}\begin{bmatrix}
        p & l & m & n & s
    \end{bmatrix}^T$, the equations for the MEE are
\begin{equation}\label{MEEdot}
    \begin{split}
        \dot{\boldsymbol z} &= {\rm \textbf{G}}\left({\boldsymbol z}, x_6\right){\boldsymbol a} \\
        \dot{x}_6 &= {\sqrt \frac{\mu}{x_1^3}} \, \eta^2 + {\sqrt \frac{x_1}{\mu}} \, \frac{x_3 \, \sin{x_6} - x_5 \, \cos{x_6}}{\eta} \, a_{h}
    \end{split}
\end{equation}
\noindent where $\mu$ represents the gravitational parameter of the main attracting body, $\eta = 1 + x_2 \cos{x_6} + x_3 \sin{x_6}$ is an auxiliary function, $\boldsymbol{a} = \begin{bmatrix} 
    a_r & a_{\theta} & a_h
\end{bmatrix}^T$ is the non-Keplerian acceleration exerted on the space vehicle projected onto $\underline{\underline{\rm R}}_{LVLH}$, and $\rm \textbf{G}$ is a $5 \times 3$ matrix depending on $\boldsymbol z$ and $x_6$ {only \cite{pontanibook}},
\begingroup
\thinmuskip=0mu
\medmuskip=0mu
\thickmuskip=0mu
\begin{equation}\label{Gmat}
    \resizebox{.88\hsize}{!}{${\rm \textbf{G}}\left({\boldsymbol z}, x_6\right)=\sqrt{\frac{x_1}{\mu}}\begin{bmatrix}
        0 & \cfrac{2x_1}{\eta} & 0 \\
        \sin{x_6} & \cfrac{(\eta+1)\cos{x_6}+x_2}{\eta} & -\cfrac{x_4\sin{x_6}-x_5 \cos{x_6}}{\eta}x_3 \\
        -\cos{x_6}
&  \cfrac{(\eta+1)\sin{x_6}+x_3}{\eta} &   \cfrac{x_4\sin{x_6}-x_5 \cos{x_6}}{\eta}x_2 \\
0 & 0 & \cfrac{1+x_4^2+x_5^2}{2\eta}\cos{x_6}\\
0 & 0 & \cfrac{1+x_4^2+x_5^2}{2\eta}\sin{x_6}
\end{bmatrix}$}
\end{equation}
\endgroup
\indent The dynamical state of the space vehicle is identified by the state vector $\boldsymbol{x}$ defined as
\begin{equation}
    \boldsymbol{x} = \begin{bmatrix}
        \boldsymbol{z}^T & x_6 & x_7
    \end{bmatrix}^T \equiv \begin{bmatrix}
        \boldsymbol{z}^T & x_6 & m_R
    \end{bmatrix}^T.
\end{equation}
\indent Two thrust angles $\alpha$ and $\beta$ define the thrust direction in terms of its components in $\underline{\underline{\rm R}}_{LVLH}$ (cf. Fig. \ref{Fig.ThrustAnglesAlphaBeta}), while the parameter $u_T$, constrained between $0$ and $u_T^{max}$, is designated to represent the thrust magnitude. Thus, the control vector is given by 
\begin{equation}
    \boldsymbol{u} = \begin{bmatrix}
        u_1 & u_2 & u_3
    \end{bmatrix}^T \equiv \begin{bmatrix}
        u_T & \alpha & \beta
    \end{bmatrix}^T
\end{equation}
\noindent and the thrust acceleration in $\underline{\underline{\rm R}}_{LVLH}$ is
\begingroup
\thinmuskip=0mu
\medmuskip=0mu
\thickmuskip=0mu
\begin{equation}
    \boldsymbol{a}_T = \begin{bmatrix} 
    a_{T,r} & a_{T,\theta} & a_{T,h}
\end{bmatrix}^T = \frac{u_1}{x_7} \begin{bmatrix}
    \rm{s}_{u_2} \rm{c}_{u_3} & \rm{c}_{u_2} \rm{c}_{u_3} & \rm{s}_{u_3}
\end{bmatrix}^T
\end{equation}
\endgroup
\noindent where $s_{\theta}$ and $c_{\theta}$ denote respectively the sine and cosine of a generic angle $\theta$.

\indent The state equations (\ref{mRdot}) and (\ref{MEEdot}) can be written in compact form as
\begin{equation}\label{dotx}
    \dot{\boldsymbol{x}} = \boldsymbol{f} \left( \boldsymbol{x}, \boldsymbol{u}, t \right). 
\end{equation}
\indent In general, the boundary conditions are problem-dependent, and are formally written in vector form as
\begin{equation}\label{zeta}
    \boldsymbol{\zeta} \left( \boldsymbol{x}_{0}, \boldsymbol{x}_{f}, t_{0}, t_{f} \right) = \begin{bmatrix}
        \boldsymbol{\zeta}_0 \left( \boldsymbol{x}_{0}, t_{0}\right)\\ \boldsymbol{\zeta}_f \left( \boldsymbol{x}_{f}, t_{f} \right)
    \end{bmatrix} = \boldsymbol 0. 
\end{equation}
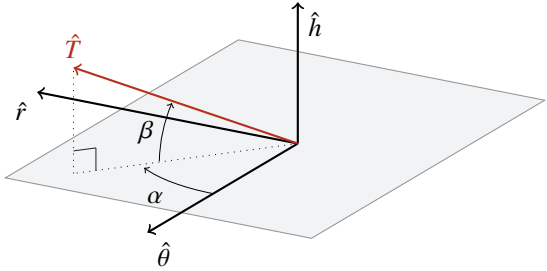
\begin{figure}[h]
    \centering
    \tdplotsetmaincoords{70}{210}
\begin{tikzpicture}[tdplot_main_coords]

\filldraw[
    draw=Gray,%
    fill=Gray!10,%
    ]          (-2,-3.1,0)
            -- (2.7,-3.1,0)
            -- (2.7,3.1,0)
            -- (-2,3.1,0)
            -- cycle;

\draw[thick,->] (0,0,0) -- (0,0,2) node[anchor=north west]{$\hat{h}$};
\draw[thick,->] (0,0,0) -- (0,4,0) node[anchor=north west]{$\hat{\theta}$};
\draw[thick,->] (0,0,0) -- (4,0,0) node[anchor=north east]{$\hat{r}$};

\pgfmathsetmacro{\ax}{2}
\pgfmathsetmacro{\ay}{2.5}
\pgfmathsetmacro{\az}{1.5}
\draw[thick,->,BrickRed] (0,0,0) -- (\ax,\ay,\az) node[anchor=south]{$\hat{T}$};


\draw[thin,dotted,Black] (0,0,0) -- (\ax,\ay,0);
\draw[thin,dotted,Black] (\ax,\ay,0) -- (\ax,\ay,\az);
\draw[very thin,Black] (\ax,\ay,0.32) -- (0.9*\ax,0.9*\ay,0.32);
\draw[very thin,Black] (0.9*\ax,0.9*\ay,0.32) -- (0.9*\ax,0.9*\ay,0);

\tdplotdrawarc[very thin,->]{(0,0,0)}{2.25}{90}{55}{anchor=north east}{$\alpha$}

\tdplotsetthetaplanecoords{53}
\tdplotdrawarc[tdplot_rotated_coords,very thin,<-]{(0,0,0)}{2}{65.5}{89}{anchor=east}{$\beta$}
 

\end{tikzpicture}
\caption{{Thrust angles $\alpha$ and $\beta$ ($0 < \alpha \leq 2\pi$, $-\nicefrac{\pi}{2} \leq \beta \leq \nicefrac{\pi}{2}$)}}
\label{Fig.ThrustAnglesAlphaBeta}
\end{figure}

\subsection{Third-body gravitational perturbation}

\indent When a spacecraft orbits a main attracting body, the gravitational action of other massive bodies, usually referred to as third bodies, can be regarded as a perturbation acting on the spacecraft. Denoting the spacecraft with $2$, the main attracting body with $1$, the third body with $3$, and neglecting the mass of the space vehicle, the gravitational perturbation exerted on body $2$ due to body $3$ is \cite{battin1999introduction} 
\begingroup
\thinmuskip=0mu
\medmuskip=0mu
\thickmuskip=0mu
\begin{equation}\label{a3B}
    \resizebox{.88\hsize}{!}{$\underrightarrow{\boldsymbol{a}}_{3B} = \mu_3\,\left\{\frac{\underrightarrow{\boldsymbol{r}}_{13} - \underrightarrow{\boldsymbol{r}}_{12}}{\left[\left( \underrightarrow{\boldsymbol{r}}_{12} -\underrightarrow{\boldsymbol{r}}_{13} \right)\cdot\left( \underrightarrow{\boldsymbol{r}}_{12} -\underrightarrow{\boldsymbol{r}}_{13} \right)\right]^{\frac{3}{2}}} - \frac{\underrightarrow{\boldsymbol{r}}_{13}}{r_{13}^3}\right\}$}
\end{equation}
\endgroup
\noindent where {$\underrightarrow{\boldsymbol{r}}_{1j}$ is the position vector of body $j$ ($=1,2$) relative to the primary, $r_{1j}$ is the position vector magnitude, and $\mu_3$ is the gravitational parameter of the perturbing body.} Table \ref{TABLE1} presents the gravitational constants and radii of the relevant celestial objects.

\begin{table}[h]
\centering
\caption{Useful planetary physical parameters}
\vspace*{-1.5mm}
\begin{tblr}{Q[c,m]|Q[c,m]|Q[c,m]}
Celestial Body & ${\mu}$  $\left[\frac{\text{km}^3}{\text{s}^2}\right]$ & $R$ $\left[\text{km}\right]$ \\
\hline 
Sun & $132712440041.279$ & not used\\
Earth & $398600.436$ & $6378.136$\\ 
Moon & $4902.800$ & $1737.400$\\
\end{tblr}
\label{TABLE1}
\end{table}

\indent Because in the governing equations for MEE the non-Keplerian acceleration appears through its components in the local vertical local horizontal reference frame, both $\underrightarrow{\boldsymbol{r}}_{12}$ and $\underrightarrow{\boldsymbol{r}}_{13}$ must be projected onto $\underline{\underline{\rm R}}_{LVLH}$, 
\begin{equation}
    \boldsymbol{r}_{12}^{(LVLH)} = \begin{bmatrix}
        r_{12} \\ 0 \\ 0
    \end{bmatrix} \;\;\;\;\;\;\;\;\; \boldsymbol{r}_{13}^{(LVLH)} = \underset{\rm LVLH \leftarrow N}{\rm \textbf{R}} \, \boldsymbol{r}_{13}^{(N)}
\end{equation}
\noindent with $\boldsymbol{r}_{13}^{(N)}$ coming from ephemeris data and N referring to either ECI or MCI. 

\section{TRAJECTORY DESIGN STRATEGY} 

\indent The investigated orbit transfer takes place in the Earth-Moon system and is inherently modeled in a multibody dynamical framework. The simultaneous gravitational action of Sun, Earth, and Moon is included along the entire transfer path. Use of MEE enables singularity-free numerical propagation and a simple writing of the boundary conditions associated with the terminal orbits. However, because during the transfer trajectory the spacecraft transitions from orbiting the Earth to orbiting the Moon, two sets of MEE must be introduced: (a) Earth-centered and ECI-related MEE ($\boldsymbol{x}_E$), associated with both $\underline{\underline{\rm N}}_{E}$ and $\underline{\underline{\rm R}}_{LVLH_E}$, and (b) Moon-centered and MCI-related MEE ($\boldsymbol{x}_M$), associated with both $\underline{\underline{\rm N}}_{M}$ and $\underline{\underline{\rm R}}_{LVLH_M}$. 

\subsection{Terminal orbits}

\indent The initial and final orbits (denoted respectively with subscripts $0$ and $f$) have specified values of some orbit elements among $a$, $e$, $i$, $\Omega$, and $\omega$. Moreover, the initial mass ratio equals $1$, meaning that $\boldsymbol{\zeta}$ has at most $11$ components (i.e., $\zeta_1$ through $\zeta_6$ associated with $\boldsymbol{\zeta}_0$ and $\zeta_7$ through $\zeta_{11}$ associated with $\boldsymbol{\zeta}_f$). The departure orbit family consists of circular low Earth orbits with specified radius $p_{E,0}$ and inclination $i_{E,0}$,
\begin{equation}\label{zeta0}
    \boldsymbol{\zeta}_0 \left( \boldsymbol{x}_{0}, t_{0} \right) = \begin{bmatrix}
       x_{1,0} - p_{E,0} \\
       x_{2,0}^2 + x_{3,0}^2\\
       x_{4,0}^2 + x_{5,0}^2 - \tan^2{\cfrac{i_{E,0}}{2}}\\
       x_{7,0} - 1
    \end{bmatrix}. 
\end{equation}
\noindent Two families of target orbits are considered in this work: (a) circular low lunar orbits with specified radius $p_{M,f}$ and inclination $i_{M,f}$,
\begin{equation}\label{1family}
    \boldsymbol{\zeta}_f \left( \boldsymbol{x}_{f}, t_{f} \right) = \begin{bmatrix}
       x_{1,f} - p_{M,f} \\
       x_{2,f}^2 + x_{3,f}^2\\
       x_{4,f}^2 + x_{5,f}^2 - \tan^2{\cfrac{i_{M,f}}{2}}
    \end{bmatrix}. 
\end{equation}
\noindent and (b) circular low lunar orbits with specified radius $p_{M,f}$, inclination $i_{M,f}$, and RAAN $\Omega_{M,f}$,
\begin{equation}\label{2family}
    \boldsymbol{\zeta}_f \left( \boldsymbol{x}_{f}, t_{f} \right) = \begin{bmatrix}
       x_{1,f} - p_{M,f} \\
       x_{2,f}^2 + x_{3,f}^2\\
       x_{4,f} - \tan{\cfrac{i_{M,f}}{2}} \cos{\Omega_{M,f}}\\
       x_{5,f} - \tan{\cfrac{i_{M,f}}{2}} \sin{\Omega_{M,f}}
    \end{bmatrix}. 
\end{equation}
\indent Terrestrial MEE are used for the initial boundary conditions associated with the departure orbit in ECI ($\boldsymbol{\zeta}_0$), while lunar MEE are used for the final boundary conditions associated with the target lunar orbit in MCI ($\boldsymbol{\zeta}_f$). 


\subsection{Trajectory arcs and coordinate transformations}\label{TACT} 

\indent During the first portion of the transfer trajectory the spacecraft orbits the Earth and its dynamical state is conveniently described with Earth-related MEE. Conversely, once the spacecraft transitions to orbital motion around the Moon, Moon-related MEE emerge as the best-suited variables for describing its dynamical state. Thus, the transfer trajectory can be virtually partitioned in two arcs: (a) a geocentric arc and (b) a selenocentric arc. The timing for transitioning from one representation to another is arbitrary, but proper selection of the instant of transition facilitates the convergence of the numerical solution process. In this study, the change in representation occurs as soon as the spacecraft enters a sphere of influence (SOI) of the Moon of radius $\rho_M$, i.e., when the following inequality is satisfied: 
\begin{equation}\label{psi}
    r_{M} - \rho_M \leq 0
\end{equation} 
\noindent where $r_{M}$ represents the instantaneous distance of the space vehicle from the Moon. Setting the SOI radius to $\rho_{M_{L_1}}$ is a convenient and reasonable choice, as $\rho_{M_{L_1}}$ corresponds to the distance of the interior libration point $L_1$ from the Moon, in the dynamical framework of the Earth-Moon CR3BP,
\begin{equation}
    \rho_{M} = \rho_{M_{L_1}} = d \left( L_1, \Moon \right)
\end{equation}
\noindent where $d \left( L_1, \Moon \right)$ indicates the distance between $L_1$ and the Moon.

\indent The procedure to obtain $\boldsymbol{x}_M$ from $\boldsymbol{x}_E$ encompasses five distinct representations for the state of the space vehicle, and is given by the following sequence of transformations: 
\begin{enumerate}[label={(\arabic*)}]
    \item\label{firsttransf} \textbf{MEE2CC}\\ transforms $\boldsymbol{x}_E = \boldsymbol{MEE}_E^{(ECI)}$ to $\boldsymbol{CC}_E^{(ECI)} = \boldsymbol{y}_E$
    
    \item\label{secondtransf} \textbf{Change of primary}\\ transforms $\boldsymbol{y}_E = \boldsymbol{CC}_E^{(ECI)}$ to $\boldsymbol{CC}_M^{(ECI)} = \boldsymbol{y}_M^{(ECI)}$ 

    
    \item\label{thirdtransf} \textbf{Change in orientation}\\ transforms $\boldsymbol{y}_M^{(ECI)} = \boldsymbol{CC}_M^{(ECI)}$ to $\boldsymbol{CC}_M^{(MCI)} = \boldsymbol{y}_M$


    

    \item\label{lasttransf} \textbf{CC2MEE}\\ transforms $\boldsymbol{y}_M = \boldsymbol{CC}_M^{(MCI)}$ to $\boldsymbol{MEE}_M^{(MCI)} = \boldsymbol{x}_M$ 
\end{enumerate}
\noindent The $i$-th step of the previous sequence, occurring at time $t_i$, can be represented through the general expression for implicit state transformations
\begin{equation} 
    \boldsymbol{\chi}_i \left( \boldsymbol{x}^{(i + 1)}, \boldsymbol{x}^{(i)}, t_i \right) = \boldsymbol 0
\end{equation}
\noindent where $\boldsymbol{x}^{(i + 1)}$ and $\boldsymbol{x}^{(i)}$ indicate the state after and before transformation $i$, respectively, and $\boldsymbol{\chi}_i$ is a nonlinear vector function, whose dimension equals the number of state components. It can be demonstrated that each individual transformation can be expressed in explicit form, which guarantees a unique solution for $\boldsymbol{x}^{(i + 1)}$ 
\begin{equation}\label{StateExplicitTransfTime} 
    \boldsymbol{x}^{(i + 1)} = {\Lambda}_i \left( \boldsymbol{x}^{(i)}, t_i \right).
\end{equation}
\noindent It is also worth mentioning that state transformations $1$, $3$, and $4$ do not explicitly depend on time.

\indent For the problem at hand, even though the trajectory design strategy requires passing through three additional representations for the dynamical state of the spacecraft, only Earth- and Moon-related MEE are used for the numerical propagation of the path of the space vehicle. Because of this, all four coordinate transformations of the sequence must be performed simultaneously, i.e., $t_1 = t_2 = t_3 = t_4$.


\section{MULTIPLE-ARC FORMULATION}\label{MAF} 


\indent This section is devoted to formulating the minimum-time transfer problem from an initial low Earth orbit to a final low lunar orbit. Low thrust is assumed to be available at any time, and the continuous thrust law associated with the optimal orbit transfers must be determined. The objective functional of the minimum-time problem can be conveniently defined in Mayer form, 
\begin{equation}\label{objfun}
    J = k_J \, t_{transf}
\end{equation}
\noindent where $k_J$ is an arbitrary positive constant and $t_{transf}$ indicates the duration of the transfer
\begin{equation}\label{ttransf}
    t_{transf} = t_{f} - t_{0}.
\end{equation}
\noindent Because $t_0$, which corresponds to the departure epoch, is specified, $t_0$ can be set to 0 in Eq. (\ref{ttransf}) and the objective functional becomes 
\begin{equation}\label{objfun1}
    J = k_J \, t_{f}.
\end{equation}

\subsection{Formulation using MEE}\label{formulationMEE}

Previous {research \cite{pontani2020optimal}} on minimum-time low-thrust orbit transfers around a single attracting body demonstrated that the use of modified equinoctial elements considerably mitigates the hypersensitivity of the numerical solution on the initial values of the time-varying adjoint variables. Because MEE are employed, the spacecraft of interest is governed by the state equations (\ref{dotx}) and is subject to boundary conditions written in the form (\ref{zeta}). In particular, the initial boundary conditions apply to the geocentric arc, while the final boundary conditions pertain to the selenocentric arc. 

\indent The state equations are directly affected by the non-Keplerian acceleration exerted on the spacecraft, i.e., the resultant of the thrust acceleration $\boldsymbol{a}_T$ and the perturbing acceleration $\boldsymbol{a}_P$, projected onto the LVLH reference frame,
\begin{equation}
    \boldsymbol{a} = \boldsymbol{a}_T + \boldsymbol{a}_P.
\end{equation}
\noindent The perturbing acceleration consists of the joint effect of the third-body gravitational perturbation exerted by the two third bodies (either Moon and Sun in the geocentric arc or Earth and Sun in the selenocentric arc). Denoting the two third bodies with subscripts $3B_1$ and $3B_2$, in general
\begin{equation}
    \boldsymbol{a}_P = \boldsymbol{a}_{3B_1} + \boldsymbol{a}_{3B_2}
\end{equation}
\noindent where the third-body perturbing acceleration is given by Eq. (\ref{a3B}).

\indent For the orbit transfer problem at hand, the set of coordinates that represent the dynamical state of the spacecraft changes as soon as the space vehicle enters the SOI of the Moon, i.e., when inequality (\ref{psi}) is fulfilled. Because the entire trajectory is composed of five arcs (three of which have zero length), this becomes a multiple-arc trajectory optimization problem with $N = 5$. In each arc $j$, the state equations are
\begin{equation}\label{dotxj}
    \dot{\boldsymbol{x}}^{(j)} = \boldsymbol{f}_j \left( \boldsymbol{x}^{(j)}, \boldsymbol{u}^{(j)}, t \right) = \boldsymbol{f}^{(j)}.
\end{equation}
\indent Denoting with $t_j$ the time at the $j$-th interface (between arc $j$ and arc $j + 1$), two additional functions are introduced
\begin{itemize}
    \item scalar transition function $\psi$, identifying the occurrence of the transition between two consecutive arcs. In general, it is a function of $\boldsymbol{x}_{ini}^{(j + 1)}$, $\boldsymbol{x}_{fin}^{(j)}$ and $t_j$,
    \begin{equation}\label{psij} 
        {\psi}_j \left( \boldsymbol{x}_{ini}^{(j + 1)}, \boldsymbol{x}_{fin}^{(j)}, t_j \right) = 0 \; 
    \end{equation}
    \item vector matching function $\boldsymbol{\chi}$, stating the (generally implicit) matching relation for the state across two adjacent arcs. In general, it is a function of $\boldsymbol{x}_{ini}^{(j + 1)}$, $\boldsymbol{x}_{fin}^{(j)}$ and $t_j$, with the same dimension as that of $\boldsymbol{x}$. The vector matching function defines an isomorphic mapping (i.e., a bijective relation) between $\boldsymbol{x}_{ini}^{(j + 1)}$ and $\boldsymbol{x}_{fin}^{(j)}$ in an open set of the state space, 
    \begin{equation}\label{chi} 
        {\boldsymbol{\chi}}_j \left( \boldsymbol{x}_{ini}^{(j + 1)}, \boldsymbol{x}_{fin}^{(j)}, t_j \right) = \boldsymbol 0.
    \end{equation}
\end{itemize}
\noindent The number of interfaces is $N - 1$ and subscripts $ini$ and $fin$ denote the initial and final value of a given variable in the arc with index reported in the superscript, respectively. 

\subsection{Necessary conditions for optimality using MEE}

\indent Optimal control theory is applied to the previously defined continuous-time dynamical system, for the purpose of obtaining the first-order necessary conditions for optimality, enabling the translation of the optimal control problem into a two-point boundary-value problem (TPBVP).

\indent As an opening step, the two building blocks of the extended objective functional, a function of boundary conditions $\Phi$ and $N$ Hamiltonian functions $H^{(j)}$, are introduced, 
\begin{equation}\label{Phix}
    \Phi = k_J \, t_{f} + \boldsymbol{\sigma}^T \boldsymbol{\zeta} + \sum\limits_{j = 1}^{N - 1} \left( \xi_j \psi_j + \boldsymbol{\nu}_j^{T} \boldsymbol{\chi}_j \right)
\end{equation}
\begin{equation}\label{Hx}
    H^{(j)} = \boldsymbol{\lambda}^{{(j)}^T} \boldsymbol{f}^{(j)}
\end{equation}
\noindent where $N$ is the total number of arcs, $\boldsymbol{\sigma}$, $\xi_j$, and $\boldsymbol{\nu}_j$ are time-independent adjoint variables conjugate to the multipoint conditions (\ref{zeta}), (\ref{psij}), and (\ref{chi}), respectively, whereas $\boldsymbol{\lambda}^{(j)}$ is the time-varying costate vector associated with the differential constraint arising from the state equations (\ref{dotxj}). The extended objective functional for multiple-arc problems is defined as
\begin{equation}
    \begin{split}
        \overline{J} &= \Phi \left( \boldsymbol{x}_{0}, \boldsymbol{x}_{f}, \boldsymbol{p}, \boldsymbol{\sigma}, t_{0}, t_{f}, t_1, \dots, t_{N - 1}, \boldsymbol{\nu}_{1}, \dots, \right. \\
        & \boldsymbol{\nu}_{N - 1}, \boldsymbol{x}_{ini}^{(2)}, \dots, \boldsymbol{x}_{ini}^{(N)}, \boldsymbol{x}_{fin}^{(1)}, \dots, \boldsymbol{x}_{fin}^{(N - 1)}, \xi_1, \dots, \\ 
        & \left. \xi_{N - 1} \right) + \sum\limits_{j = 1}^{N} \int_{t_{j - 1}}^{t_{j}} \left[ H^{(j)} \left( \boldsymbol{x}^{(j)}, \boldsymbol{u}^{(j)}, \boldsymbol{p}, \boldsymbol{\lambda}^{(j)}, t \right) + \right. \\
        & \left.- \boldsymbol{\lambda}^{{(j)}^T} \dot{\boldsymbol{x}}^{(j)} \right] dt
    \end{split}
\end{equation}
\noindent where $t_N = t_f$.

\indent The first differential of the augmented objective functional ${\rm d} \overline{J}$ can be obtained using the chain rule after lengthy {developments \cite{pontani2021optimal}}, omitted for the sake of brevity. Then, the necessary conditions for optimality can be derived by requiring ${\rm d} \overline{J} = 0$ 
\begin{equation}\label{ext1} 
    \frac{\partial \Phi}{\partial \boldsymbol{x}_{ini}^{(j + 1)}} = -\boldsymbol{\lambda}_{ini}^{{{(j + 1)}}^T}
\end{equation}
\begin{equation}\label{ext2}
    \frac{\partial \Phi}{\partial \boldsymbol{x}_{fin}^{(j)}} = \boldsymbol{\lambda}_{fin}^{{(j)}^T}
\end{equation}
\begin{equation}\label{ext3}
    \frac{\partial \Phi}{\partial t_j} = H_{ini}^{(j + 1)} - H_{fin}^{(j)}
\end{equation}
\begin{equation}\label{lambdabnd}
        \boldsymbol{\lambda}_{{0}} = -\left( \frac{\partial \Phi}{\partial \boldsymbol{x}_{0}} \right)^T \;\;\;\;\;\;\;\;\; \boldsymbol{\lambda}_{{f}} = \left( \frac{\partial \Phi}{\partial \boldsymbol{x}_{f}} \right)^T
\end{equation}
\begin{equation}\label{Htransv}
    H_{0} = \frac{\partial \Phi}{\partial t_{0}} \;\;\;\;\;\;\;\;\;\;\;\;\;\; H_{f} = -\frac{\partial \Phi}{\partial t_{f}}
\end{equation}
\begin{equation}\label{ELnecconds1MA}
    \dot{\boldsymbol{\lambda}}^{(j)} = -\left( \frac{\partial H^{(j)}}{\partial \boldsymbol{x}^{(j)}} \right)^T
\end{equation}
\begin{equation}\label{Pontry}
    \boldsymbol{u}_*^{(j)} = \underset{{\; \; \; \; \; \; \; \; \boldsymbol{u}^{(j)}}}{\rm arg \, min} \, H^{(j)}
\end{equation}
\begin{equation}\label{pnecconds}
        \sum\limits_{j = 1}^{N} \int_{t_{j - 1}}^{t_{j}} \frac{\partial H^{(j)}}{\partial \boldsymbol{p}} {\rm d}t + \frac{\partial \Phi}{\partial \boldsymbol{p}} = \boldsymbol{0}^T
\end{equation}
\noindent where subscript $*$ denotes the optimal value of the corresponding variable {and $\boldsymbol p$ is a parameter vector collecting time-independent parameters which may be needed in modeling the dynamical system.} 

\indent Application of the Pontryagin minimum principle (\ref{Pontry}) yields the optimal expressions for the control components,
\begin{equation}\label{controlxoptj}
    \begin{split}
        u_{1}^{(j,opt)} &= u_T^{(max)}\\
        u_{2}^{(j,opt)} &= 2 \, \arctan{\frac{-H_{r}^{(j)}}{\sqrt{H_{r}^{{(j)}^2} + H_{{\theta}}^{{(j)}^2}} - H_{{\theta}}^{(j)}}}\\
        u_{3}^{(j,opt)} &= \arcsin{\frac{-H_{h}^{{(j)}}}{\sqrt{H_{r}^{{(j)}^2} + H_{{\theta}}^{{(j)}^2} + H_{h}^{{(j)}^2}}}}
    \end{split}
\end{equation}
\noindent where $H_{r}^{{(j)}}$, $H_{{\theta}}^{{(j)}}$, and $H_{h}^{{(j)}}$ are the coefficients that multiply $a_{T,r}^{{(j)}}$, $a_{{T,\theta}}^{{(j)}}$, and $a_{T,h}^{{(j)}}$ in the expression for $H^{(j)}$, respectively.

\indent The consequence of maximum thrusting at all times is that the mass depletion rate becomes constant and the time behavior of the mass ratio can be described by the following analytical expression:
\begin{equation}\label{x7anal}
    m_R \left( t \right) = 1 - \frac{u_T^{(max)}}{c} \, t.
\end{equation}
\noindent This allows neglecting the mass ratio as a state component and $u_{1}$ as a control component in the optimization process. Moreover, the optimal expressions for the control components lead to the following formulation for the Hamiltonian:
\begin{equation}\label{Hxoptj}
    H^{(j,opt)} = H_{1}^{{(j)}} - \frac{u_T^{(max)}}{m_R} \sqrt{H_{r}^{{(j)}^2} + H_{{\theta}}^{{(j)}^2} + H_{h}^{{(j)}^2}}
\end{equation}
\noindent where $H_{1}^{{(j)}}$ is the portion of $H^{(j)}$ independent of $u_1^{{(j)}}$. The expressions for $H_{1}^{{(j)}}$, $H_{r}^{{(j)}}$, $H_{{\theta}}^{{(j)}}$, and $H_{h}^{{(j)}}$, obtained after lengthy developments, depend on both $\boldsymbol{x}^{(j)}$ and $\boldsymbol{\lambda}^{(j)}$. Neglecting all superscripts for the sake of conciseness, the Hamiltonian can be rewritten as
\begin{equation}\label{HxScalCostate}
    \begin{split}
        H^{(opt)} =& \overbrace{H_{1,0} + H_{r}  a_{P,r} + H_{{\theta}}  a_{P,\theta} + H_{h}  a_{P,h}}^{H_{1}} +\\
        -&\frac{u_T^{(max)}}{m_R} \sqrt{H_{r}^2 + H_{{\theta}}^2 + H_{h}^2}
    \end{split}.
\end{equation}
\noindent The blocks $H_{1,0}$, $H_{r}$, $H_{{\theta}}$ and $H_{h}$ in Eq. (\ref{HxScalCostate}) are all linear in $\boldsymbol{\lambda}$
\begin{align}
    H_{1,0} &= \lambda_{6}  \sqrt{\frac{\mu}{x_1^3}} \left( 1 + x_2  \cos{x_6} + x_3  \sin{x_6} \right)^2\\
    H_{{r}} &= \sqrt{\frac{x_1}{\mu}}  \left( \lambda_{2}  \sin{x_6} - \lambda_{3}  \cos{x_6} \right)
\end{align}
\begin{align}
    \begin{split}
        H_{{\theta}} &= \frac{\sqrt{\frac{x_1}{\mu}}}{1 + x_2  \cos{x_6} + x_3  \sin{x_6}}  \left\{ 2  x_1  \lambda_{1} + \right.\\
        &+ \left[x_3 + \cos{x_6}  \left(2 + x_2  \cos{x_6} + x_3  \sin{x_6} \right) \right]  \lambda_{2} +\\
        &+ \left. \left[ x_3 + \sin{x_6}  \left(2 + x_2  \cos{x_6} + x_3  \sin{x_6} \right) \right]  \lambda_{3} \right\}
    \end{split}\\
    \begin{split}
        H_{{h}} &= \frac{\sqrt{\frac{x_1}{\mu}}}{2  \left( 1 + x_2  \cos{x_6} + x_3  \sin{x_6} \right)}  \left[  2  \left( x_4  \sin{x_6} + \right. \right.\\
        &- \left. x_5  \cos{x_6} \right)  \left( \lambda_{6} + x_2  \lambda_{3} - x_3  \lambda_{2} \right) +\\
        &+ \left. \left( x_4^2 + x_5^2 + 1 \right)  \left( \cos{x_6}  \lambda_{4} + \sin{x_6}  \lambda_{5} \right) \right].
    \end{split}
\end{align}
\noindent Linearity implies homogeneity of the costate equations. In fact, assuming $\tilde{\boldsymbol{\lambda}} = k_{\lambda}  \boldsymbol{\lambda}$ would yield $\tilde{H} = k_{\lambda} H$, and ultimately
\begin{equation}\label{CostateHomogeneity}
    \dot{\tilde{\boldsymbol{\lambda}}} = \frac{\partial \tilde{H}}{\partial \boldsymbol{x}} = k_{\lambda}  \frac{\partial H}{\partial \boldsymbol{x}} = k_{\lambda}  \dot{\boldsymbol{\lambda}}
\end{equation}
\noindent where $k_{\lambda}$ is an arbitrary positive constant. As a result, although the costate components have no upper or lower bound, the search space for the costate variables can be reduced to an arbitrary interval $\left[ \lambda_{min}, \; \lambda_{max} \right]$, provided that $\lambda_{min} < 0$ and $\lambda_{max} > 0$. This property, i.e., \textit{scalability of the costate}, can be leveraged to restrict the search space explored by the numerical solution process.

\indent The initial transversality condition on the Hamiltonian from Eq. (\ref{Htransv}) must not be enforced because $t_{0}$ is specified and set to zero. On the other hand, the final transversality condition on the Hamiltonian yields
\begin{equation}
    H_{{f}} = -\frac{\partial \Phi}{\partial t_{f}} = -k_J.
\end{equation}
\noindent Since $k_J$ is an arbitrary positive constant, the final transversality condition is equivalent to the following inequality constraint:
\begin{equation}\label{neccondsHxfin}
    H_{{f}} < 0.
\end{equation}
\subsection{Implicit costate transformation}

\indent The formulation of the Earth-Moon orbit transfer as a multiple-arc optimization problem leads to establishing an extended set of necessary conditions for optimality. In particular, the multipoint necessary conditions for optimality (\ref{ext1})-(\ref{ext3}), together with the multipoint conditions (\ref{psij}) and (\ref{chi}), represent a considerable number of additional relations inherent to the multiple-arc formulation. These additional conditions must be enforced in the numerical solution process. However, Eqs. (\ref{ext1})-(\ref{ext3}) can be combined, for the purpose of obtaining the matching relation for the costate variables between two consecutive arcs in an advantageous form. {Appendix A proves that, for the problem at hand, the costate jump relation to enforce at the transitions corresponding to each coordinate transformation $i$ ($=1,...,4$) can be obtained via implicit costate transformation,}
\begin{equation}\label{CostateImplicit}
        \left( \frac{\partial \boldsymbol{\chi}_j}{\partial \boldsymbol{x}_{ini}^{(j + 1)}} \right)^{-T} \boldsymbol{\lambda}_{ini}^{{(j + 1)}} + \left( \frac{\partial \boldsymbol{\chi}_j}{\partial \boldsymbol{x}_{fin}^{(j)}} \right)^{-T} \boldsymbol{\lambda}_{fin}^{{(j)}} = 0.
\end{equation}
\noindent This single relation can be manipulated to obtain closed-form expressions for either $\boldsymbol{\lambda}_{ini}^{{(j + 1)}}$ or $\boldsymbol{\lambda}_{fin}^{{(j)}}$. Both relations are critical to the numerical solution process, whose algorithmic steps are described in Section \ref{MS}.

\subsection{Initial conditions for the geocentric arc} 

\indent The initial conditions pertain to the geocentric leg, where Earth-centered and ECI-related MEE are employed. Application of Eq. (\ref{lambdabnd}) to the initial conditions (\ref{zeta0}) yields the following conditions on the initial costate:
\begin{equation}\label{lambdabnd0}
    \begin{split}
        \lambda_{{{2,0}}} \, x_{{3,0}} - \lambda_{{{3,0}}} \, x_{{2,0}} &= 0\\
        \lambda_{{{4,0}}} \, x_{{5,0}} - \lambda_{{{5,0}}} \, x_{{4,0}} &= 0\\
        \lambda_{{{6,0}}} &= 0
    \end{split}.
\end{equation}
\subsection{Final conditions for the selenocentric arc} 

\indent The final conditions pertain to the selenocentric leg, where Moon-centered and MCI-related MEE are employed. Application of Eq. (\ref{lambdabnd}) to the final conditions (\ref{1family}), with free RAAN, and (\ref{2family}), with specified RAAN, yields the following conditions on the final costate, respectively: 
\begin{equation}\label{lambdabndf1}
    \begin{split}
        \lambda_{{{2,f}}} \, x_{{3,f}} - \lambda_{{{3,f}}} \, x_{{2,f}} &= 0\\
        \lambda_{{{4,f}}} \, x_{{5,f}} - \lambda_{{{5,f}}} \, x_{{4,f}} &= 0\\
        \lambda_{{{6,f}}} &= 0
    \end{split}    
\end{equation}
\begin{equation}\label{lambdabndf2}
    \begin{split}
        \lambda_{{{2,f}}} \, x_{{3,f}} - \lambda_{{{3,f}}} \, x_{{2,f}} &= 0\\
        \lambda_{{{6,f}}} &= 0
    \end{split}.
\end{equation}

\section{METHOD OF SOLUTION}\label{MS} 

This section focuses on the numerical solution strategy. The technique being described aims at fulfilling all the necessary conditions arising from the translation of the multiple-arc optimal control problem into a TPBVP. The indirect approach is pursued via numerical search over the solution space, performed through a heuristic optimization algorithm.

\subsection{Stratified-objective strategy}\label{ProObjStr} 

 The heuristic optimization algorithm explores the search space and follows a distinctive procedural behavior aimed at locating the minimum of a given scalar fitness function $\Gamma$. When ordinary optimal control problems are tackled, a fitness function is introduced that evaluates the boundary conditions violation, for a specific choice of the parameter set. On the other hand, problems of greater complexity call for a shrewder fitness function design. In this research, because the primary changes along the orbit transfer, some intermediate objectives are introduced, besides the satisfaction of the necessary conditions at the boundaries, for the purpose of facilitating the convergence of the numerical solution process. In simpler terms, the fitness function is structured in a way that allows using different expressions according to the performance of the solution at hand.

\indent Two intermediate objectives ($\Gamma_1$ and $\Gamma_2$) are designed in order to prompt two consecutive key events ($E_{f_{1}}$ and $E_{f_{2}}$) that must occur in every Earth-Moon orbit transfer: 
\begin{enumerate}[label={($E_{f_{\arabic*}}$)}]
    \item\label{obj1} the spacecraft enters the SOI of the Moon; 
    \item\label{obj2} the Moon captures the spacecraft (i.e., the specific orbital energy of the spacecraft relative to the Moon drops below a preselected fraction of that of the target orbit).
\end{enumerate}



\noindent These two events represent the two separations between the three possible definitions, or layers, for $\Gamma$
\begin{equation}\label{FitFun}
    \Gamma \left( \boldsymbol{\Upsilon} \right) = \begin{cases}
        \Gamma_1 \left( \boldsymbol{\Upsilon} \right) \; \text{before} \; E_{f_1}\\
        \Gamma_2 \left( \boldsymbol{\Upsilon} \right) \; \text{between} \; E_{f_1} \; \text{and} \; E_{f_2}\\
        \Gamma_3 \left( \boldsymbol{\Upsilon} \right) \; \text{after} \; E_{f_2}
    \end{cases}
\end{equation}
\noindent where $\boldsymbol{\Upsilon}$ denotes the input parameter set in vector form.


\indent Each formulation is aimed at shepherding the population members to the next level. In fact, $\Gamma_1$ and $\Gamma_2$ are proportional to the spacecraft-Moon distance and to the specific orbital energy of the spacecraft relative to the Moon, respectively, whereas $\Gamma_3$ retains the typical formulation associated with ordinary optimal control problems. Furthermore, scaling constants are introduced so that the algorithm assigns a lower cost to a specific solution as it evolves to the next objective.

\subsection{Indirect heuristic algorithm}\label{IHA}

The problem is solved with an indirect heuristic approach, by enforcing all the necessary conditions for optimality, i.e., by solving the corresponding TPBVP. In this work, the heuristic algorithm is differential {evolution\cite{storn1997differential}}. The indirect heuristic method has three major favorable features, i.e. (i) absence of any starting guess, because only the search space for the unknown parameters is to be specified, (ii) no special representation is assumed for the control variables time histories, and (iii) enforcement of the analytical conditions for optimality is obtained at the end of the process.

\indent The numerical solution process employs a population of individuals, where each population member corresponds to a unique set of unknown parameters. A detailed compendium of the key steps of the solution method is provided in the following algorithm: \\
\\
\noindent \textbf{Algorithm 1}: Forward Indirect Heuristic Algorithm

\begin{enumerate}
    \vspace{-0.2cm}\item Identify the minimum set of parameters required to compute the unknown components of the state and costate vectors $\boldsymbol{x}_{E,0}$ and $\boldsymbol{\lambda}_{E,0}$ at $t_0$. 
    \vspace{-0.2cm}\item For each individual $i$, with $i \leq N_P$ (with $N_P$ denoting the number of individuals), iterate the following sub-steps:
    \begin{enumerate}[label=\small{\alph*}.]
            
            \vspace{-0.2cm}\item\label{FirstSubStepF} select the input values for the minimum set of parameters identified at step 1; 

           \item compute the remaining unknown components of the state and costate vectors $\boldsymbol{x}_{E,0}$ and $\boldsymbol{\lambda}_{{E,0}}$ from the minimum set of parameters using the necessary conditions at $t_0$; 
             
            \item select the time of flight $t_{f}$ and set the mass ratio at $t_0$ to $m_{R,0} = 1$;

            item forward propagate both state and costate in the geocentric arc with the respective governing equations, using the expressions presented in Eq. (\ref{controlxoptj}) for the control components, until either an event condition is met or $t = t_{f}$. If event $E_{f_1}$ does not occur, then compute $\Gamma_1$, and consider the next individual (i.e., go to step 2);

            \item apply the interface conditions in order to obtain the values for the state and costate vectors $\boldsymbol{x}_{M,ini}$ and $\boldsymbol{\lambda}_{{M,ini}}$ (related to the selenocentric segment) from the values of the state and costate vectors $\boldsymbol{x}_{E,fin}$ and $\boldsymbol{\lambda}_{{E,fin}}$ (related to the geocentric segment) at the interface time $t_1$, following the procedure described in Section \ref{TACT}; 

            \item forward propagate both state and costate in the selenocentric arc with the respective governing equations, using the expressions presented in Eq. (\ref{controlxoptj}) for the control components, until either an event condition is met or $t = t_{f}$. If event $E_{f_2}$ does not occur, then compute $\Gamma_2$, and consider the next individual; 

           \item continue the propagation of the selenocentric arc until either an event condition is met or $t = t_{f}$; 
            
           \item\label{hF} evaluate the Hamiltonian at $t_f$, $H_{f}$. If $H_{f} < 0$ proceed to the next step, otherwise set the value of $\Gamma_3$ to the minimum allowed value of $\Gamma_2$ and consider the next individual;
            
            \item evaluate the violation of the necessary conditions at $t_f$ and then compute $\Gamma_3$, whose value is a measure of the violation of the final equality constraints. Then, consider the next individual.

        \end{enumerate}
    \vspace{-0.2cm}\item  Once the fitness functions are obtained for all the population members, use the differential evolution technique to generate a new population. 
    \vspace{-0.2cm}\item Iterate the previous steps until some stopping criterion is met (either the fitness function value associated with the best individual drops below a given threshold or the maximum number of iteration is reached).
\end{enumerate}
 
\vspace{-0.2cm}\indent In the case of free final RAAN, Algorithm 1 was used. This version of the indirect heuristic algorithm is the most straightforward, mainly because it is based on forward propagation. However, in solving TPBVP, propagation can be performed either forward or backward in time, as long as the differential constraints are held throughout the entire path.

\indent A similar version of the algorithm, based on backward propagation, was used in the case of specified final RAAN. Backward propagation is made possible thanks to the existence of the analytical expression for the mass ratio (\ref{x7anal}). This allows the identification of a minimum set of parameters for the backward propagation algorithm that retains the same size as that used in the forward propagation algorithm. Backward propagation is advantageous in the case of specified final RAAN, because the final orbit family is more constrained than the initial orbit family. The backward algorithm is undoubtedly less intuitive with respect to the forward version. Furthermore, the two consecutive key events ($E_{b_{1}}$ and $E_{b_{2}}$) in the backward case are not only the opposite of those presented in Section \ref{ProObjStr}, but are also in reverse order:
\begin{enumerate}[label={($E_{b_{\arabic*}}$)}]
    \vspace{-0.2cm}\item the spacecraft escapes the Moon (i.e., the specific orbital energy of the spacecraft relative to the Moon rises above a preselected fraction of that of the target orbit);
    \vspace{-0.2cm}\item the spacecraft leaves the SOI of the Moon. 
\end{enumerate} 
\vspace{-0.2cm}\noindent In this case, subscripts $ini$ and $fin$ refer to the new independent variable $\tau = t_f - t$. Therefore, the matching conditions at the interface relate ($x_{E,ini}^{(j+1)}$, $\lambda_{E,ini}^{(j+1)}$) to ($x_{M,fin}^{(j)}$, $\lambda_{M,fin}^{(j)}$), because the geocentric arc is encountered after the selenocentric arc in the backward integration process. \\
\\
\noindent \textbf{Algorithm 2}: Backward Indirect Heuristic Algorithm

\begin{enumerate}
    \vspace{-0.2cm}\item Identify the minimum set of parameters required to compute the unknown components of the state and costate vectors $\boldsymbol{x}_{M,f}$ and $\boldsymbol{\lambda}_{M,f}$ at $t_f$. 
    \item For each individual $i$, with $i \leq N_P$ (with $N_P$ denoting the number of individuals), iterate the following sub-steps:
    \begin{enumerate}[label=\small{\alph*}.]

            \vspace{-0.2cm}\item\label{FirstSubStepB} select the input values for the minimum set of parameters identified at step 1; 

            \item compute the remaining unknown components of the state and costate vectors $\boldsymbol{x}_{M,f}$ and $\boldsymbol{\lambda}_{{M,f}}$ from the minimum set of parameters using the necessary conditions at $t_f$; 

            \item evaluate the Hamiltonian at $t_f$, $H_{f}$. If $H_{f} \geq 0$ set the value of $\Gamma_1$ to infinity and consider the next individual (i.e., go to step 2);

           \item select the time of flight $t_{f}$ and set the mass ratio at $t_{f}$ to $m_{R,f} = 1 - \nicefrac{t_{f} u_T^{(max)}}{c}$;

           \item backward propagate both state and costate in the selenocentric arc with the respective governing equations, using the expressions presented in Eq. (\ref{controlxoptj}) for the control components, until either an event condition is met or $t = t_0$. If event $E_{b_1}$ does not occur, then compute $\Gamma_1$, and consider the next individual; 

            \item continue the propagation of the selenocentric arc until either an event condition is met or $t = t_0$. If event $E_{b_2}$ does not occur, then compute $\Gamma_2$, and consider the next individual; 

           \item apply the interface conditions in order to obtain the values for the state and costate vectors $\boldsymbol{x}_{E,ini}$ and $\boldsymbol{\lambda}_{{E,ini}}$ (related to the geocentric segment) from the values of the state and costate vectors $\boldsymbol{x}_{M,fin}$ and $\boldsymbol{\lambda}_{{M,fin}}$ (related to the selenocentric segment) at the interface time $t_1$, following the inverse procedure of that needed in Algorithm 1; 

            \item backward propagate both state and costate in the geocentric arc with the respective governing equations, using the expressions presented in Eq. (\ref{controlxoptj}) for the control components, until either an event condition is met or $t = t_{0}$;

            \item evaluate the violation of the necessary conditions at $t_0$ and then compute $\Gamma_3$, whose value is a measure of the violation of the initial equality constraints. Then, consider the next individual.
        \end{enumerate}

        \vspace{-0.2cm}\item Once the fitness functions are obtained for all the population members, use the differential evolution technique to generate a new population. 
        \vspace{-0.2cm}\item Iterate the previous steps until some stopping criterion is met (either the fitness function value associated with the best individual drops below a given threshold or the maximum number of iteration is reached).
        
\end{enumerate}

\indent The performance of the numerical implementation of the previously described solution process can be improved by introducing canonical units, for both distance and time. The distance unit DU and time unit TU are chosen such that
\begin{equation}\label{DUTUgeneral}
    \text{DU} = R_E \;\;\;\;\;\;\;\;\; \mu_E = 1 \; \frac{\text{DU}^3}{\text{TU}^2}.
\end{equation}
\section{NUMERICAL RESULTS} 

\indent A low-thrust three-dimensional orbit transfer is considered in the non-autonomous continuous-time Sun-Earth-Moon dynamical framework. The space vehicle is modeled as a point mass equipped with a constant-specific-impulse and thrust-limited engine, whose performance is characterized by the two following propulsion parameters, chosen in compliance with current technological capabilities: 
\begin{equation}
        u_T^{(max)} = 10^{-4} \, g_0 \;\;\;\;\;\;\;\;\;
        c = 30 \; \cfrac{\text{km}}{\text{s}}
\end{equation}
\noindent with $g_0 =  9.80665 \; \nicefrac{\text{m}}{\text{s}^2}$. The departure epoch, which determines the positions of the three mentioned celestial bodies at the beginning of the orbit transfer, is set to June 1, 2025 at 0:00 UTC. 










\indent The departure low Earth orbit is selected to be circular, with specified altitude and inclination 
\begin{equation}\label{OrbitIniE} 
        p_{E,0} = R_E + 452 \; \text{km} \;\;\;\;\;\;\;\;\;
        i_{E,0} = 51.6 \; \text{\degree}.
\end{equation}
\noindent The target low lunar orbit is selected to be circular, with specified altitude and inclination 
\begin{equation}\label{OrbitFinM}
        p_{M,f} = R_M + 100 \; \text{km} \;\;\;\;\;\;\;\;\;
        i_{M,f} = 90 \; \text{\degree}.
\end{equation}
\noindent The choice of a family of low-altitude circular polar lunar orbit is convenient for a wide range of applications of practical interest. First of all, coverage of the lunar south pole and easy access to the lunar surface. 

\subsection{Free final RAAN}

\indent As a first step, the minimum set of parameters associated with a solution to the problem at hand is obtained, together with each parameter search interval (corresponding to step 1 of Algorithm 1). The minimum set of parameters is retrieved after introducing the time-independent parameters $\Omega_{E,0}$, that is the initial value of the RAAN, and $\lambda_{{E_{4,5,0}}}$, a useful parameter for the initial values of the fourth and fifth component of the costate vector.

\indent The initial state and costate vectors are expressed as functions of the previously introduced parameters as
\begingroup
\thinmuskip=0mu
\medmuskip=0mu
\thickmuskip=0mu
\begin{equation}\label{xE0} 
\resizebox{.88\hsize}{!}{$\begin{split}
    \boldsymbol{x}_{E,0} &= \begin{bmatrix}
        p_{E,0} & 0 & 0 & \tan{\frac{i_{E,0}}{2}} c_{\Omega_{E,0}} & \tan{\frac{i_{E,0}}{2}} s_{\Omega_{E,0}} & x_{E_{6,0}}
    \end{bmatrix}^T \\
    \boldsymbol{\lambda}_{{E,0}} &= \begin{bmatrix}
        \lambda_{{E_{1,0}}} & \lambda_{{E_{2,0}}} & \lambda_{{E_{3,0}}} & \lambda_{{E_{4,5,0}}} c_{\Omega_{E,0}} & \lambda_{{E_{4,5,0}}} s_{\Omega_{E,0}} & 0
    \end{bmatrix}^T.
\end{split}$}
\end{equation}
\endgroup
\noindent It is easy to verify that this peculiar representation for the initial state and costate vectors satisfies the necessary conditions at the initial time (\ref{zeta0}) and (\ref{lambdabnd0}), and enables the definition of the minimum set of initial parameters 
\begin{equation}\label{MinParSet}
    \resizebox{.88\hsize}{!}{$\boldsymbol{\Upsilon} = \begin{bmatrix}
        t_{f} & \Omega_{E,0} & x_{E_{6,0}} & \lambda_{{E_{1,0}}} & \lambda_{{E_{2,0}}} & \lambda_{{E_{3,0}}} & \lambda_{{E_{4,5,0}}}
    \end{bmatrix}^T.$}
\end{equation}
\noindent Each individual corresponds to a specific selection of these unknown parameters. The lower and upper bounds of the search space of each parameter are collected in two vectors, 
\begin{equation}\label{LBUBFreeRAAN}
\resizebox{.88\hsize}{!}{$\begin{split}
    \boldsymbol{LB}_{\Upsilon} &= \begin{bmatrix}
      t_{f}^{(min)} & -\pi & -\pi & \lambda_{min} & \lambda_{min} & \lambda_{min} & \lambda_{min}
    \end{bmatrix}^T \\ 
    \boldsymbol{UB}_{\Upsilon} &= \begin{bmatrix}
      t_{f}^{(max)} & \;\;\pi & \phantom{-}\pi & \lambda_{max} & \lambda_{max} & \lambda_{max} & \lambda_{max} \end{bmatrix}^T
\end{split}$}
\end{equation}
\noindent where $t_{f}^{(min)} = \tau \, \hat{t}_{f}$ and $t_{f}^{(max)} = \nicefrac{\hat{t}_{f}}{\tau}$, $\hat{t}_{f}$ is a first-order estimate of the transfer time, whose derivation is provided in Appendix B, whereas $\tau = \nicefrac{3}{4}$ is a positive constant, whose value indicates the confidence on the estimate of the transfer time. Furthermore, $\lambda_{min} = -1$ and $\lambda_{max} = 1$, in compliance with the scalability of the costate variables discussed in Section \ref{formulationMEE}.


\indent The equality necessary conditions at the final time (\ref{1family}) and (\ref{lambdabndf1}) are collected in the auxiliary vector $\boldsymbol{Q}$, 
\begin{equation}\label{Q}
    \boldsymbol{Q} = \begin{bmatrix}
        x_{M_{1,f}} - p_{M,f} \\ x_{M_{2,f}}^2 + x_{M_{3,f}}^2 - e_{M,f}^2 \\ x_{M_{4,f}}^2 + x_{M_{5,f}}^2 - \tan^2{\cfrac{i_{M,f}}{2}} \\ \lambda_{{M_{2,f}}} \, x_{M_{3,f}} - \lambda_{{M_{3,f}}} \, x_{M_{2,f}} \\ \lambda_{{M_{4,f}}} \, x_{M_{5,f}} - \lambda_{{M_{5,f}}} \, x_{M_{4,f}} \\ \lambda_{{M_{6,f}}}
    \end{bmatrix}.
\end{equation}
\noindent Thus, fitness function $\Gamma_3$ employed by the forward numerical solution method is
\begin{equation}\label{Gamma3F}
    \resizebox{.88\hsize}{!}{$\Gamma_3 \left( \boldsymbol{\Upsilon} \right) = \sqrt{w_1 Q_1^2 + w_2 Q_2^2 + w_3 Q_3^2 + w_4 Q_4^2 + w_5 Q_5^2 + w_6 Q_6^2}$}
\end{equation}
\noindent where $Q_i$ (with $i = 1, \dots, 6$) is the 
$i$-th component of vector $\boldsymbol{Q}$, and the weighting coefficients $w_i$ are all set to 1, except for $w_2 = 100$.

\indent The numerical solution process employs 200 individuals. After 2956 iterations, the best individual corresponds to a value of the fitness function equal to $4.018 \cdot 10^{-7}$. The time of flight equals 84.85 days (geocentric and selenocentric arcs lasting 71.84 days and 13.01 days, respectively), whereas the final mass ratio is 0.7604. The optimal initial and final RAAN equal $-22.5$ \degree and $207.3$ \degree, respectively. 

\begin{figure}[H]
    \centering
    \includegraphics[width=0.49\textwidth]{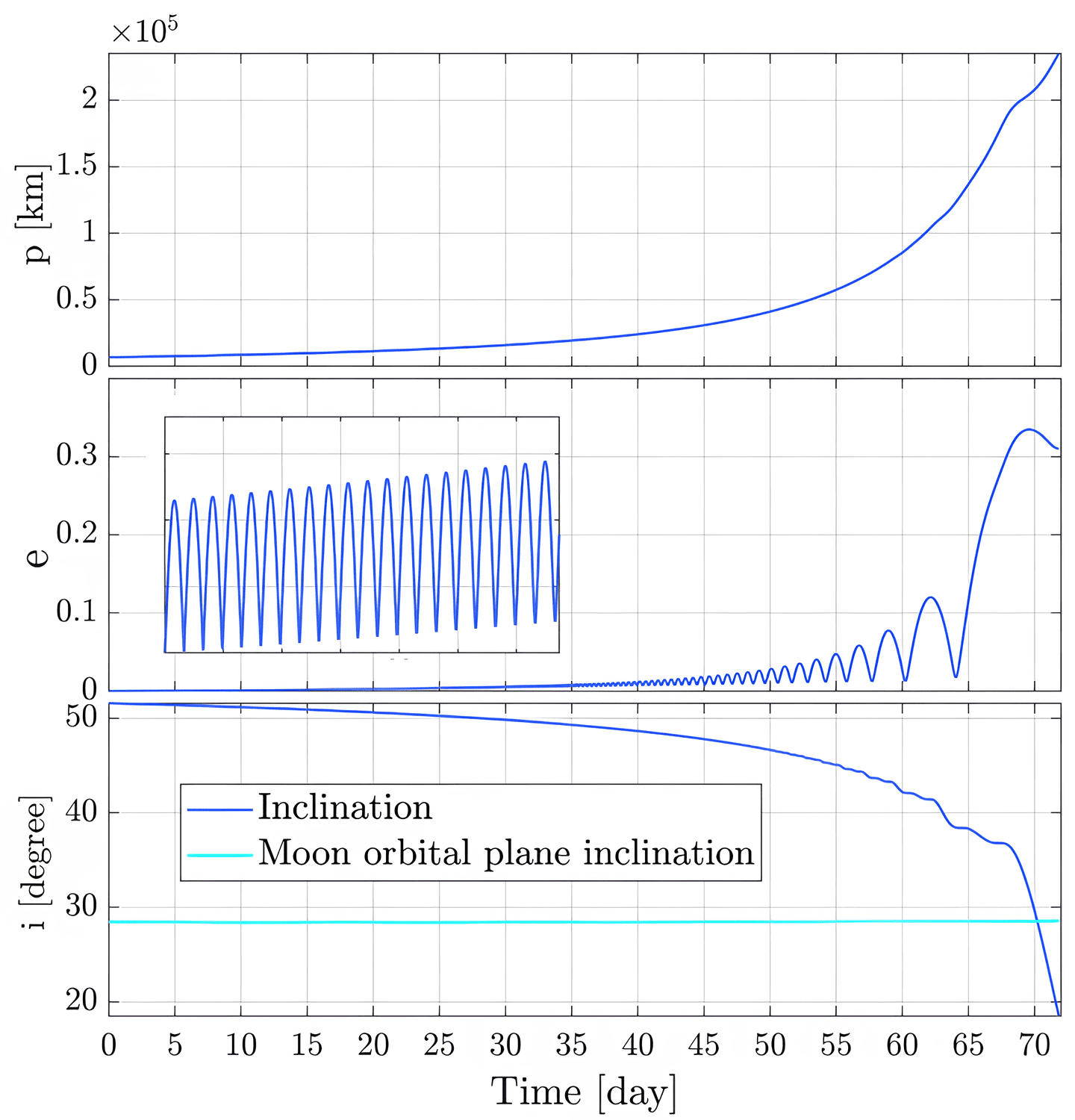}
    \vspace*{-6.9mm}
    \caption{Time histories of $p$, $e$, and $i$ in the geocentric leg}
    \label{Fig.peiE}
\end{figure}

\begin{figure}[H]
    \centering
    \includegraphics[width=0.49\textwidth]{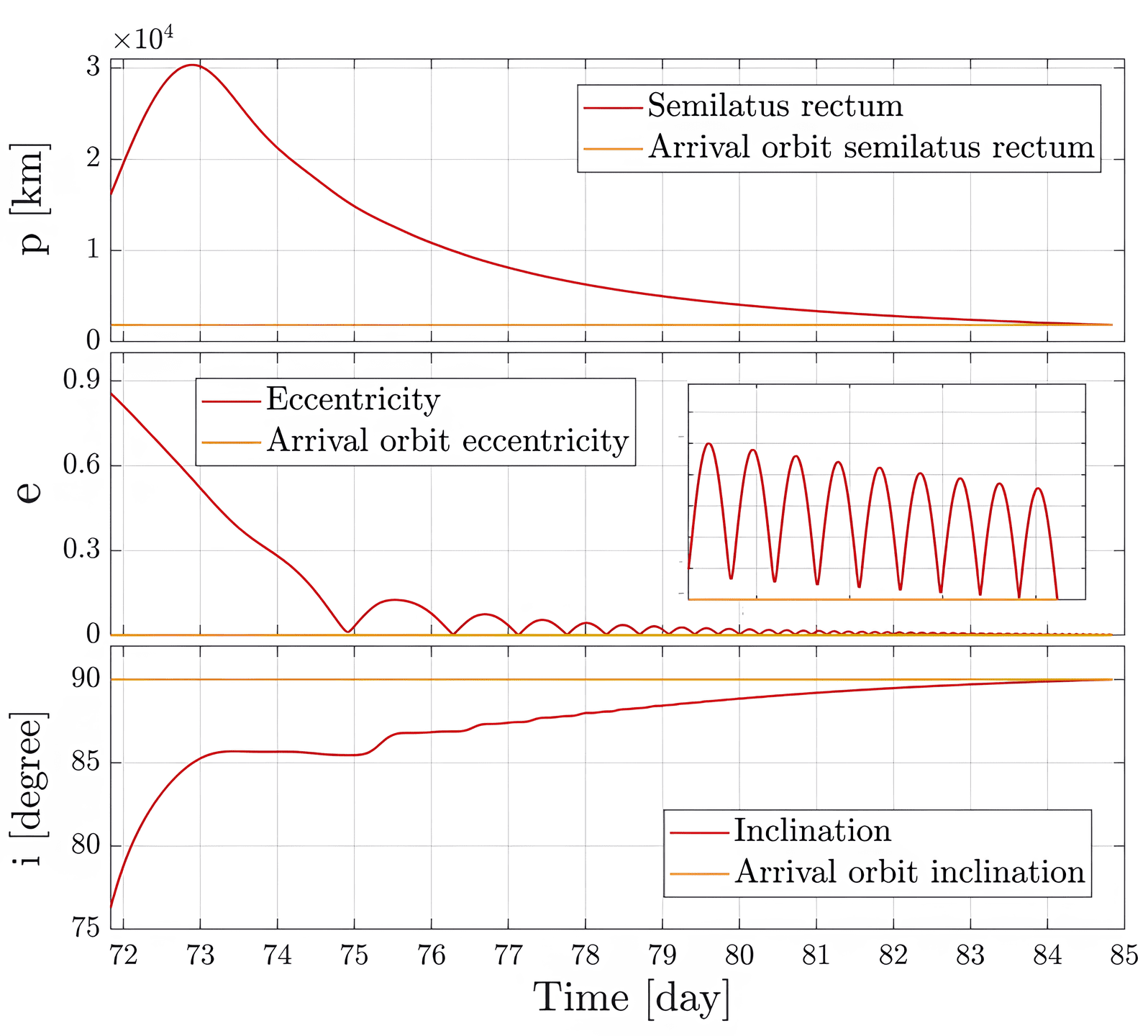}
    \vspace*{-6.9mm}
    \caption{Time histories of $p$, $e$, and $i$ in the selenocentric leg}
    \label{Fig.peiM}
\end{figure}

\indent Figure \ref{Fig.peiE} depicts the time histories of Earth-centered and ECI-related $p$, $e$, and $i$ in the geocentric leg, while Fig. \ref{Fig.peiM} depicts the time histories of Moon-centered and MCI-related $p$, $e$, and $i$ in the selenocentric leg.

\begin{figure}
    \centering
    \includegraphics[width=0.49\textwidth]{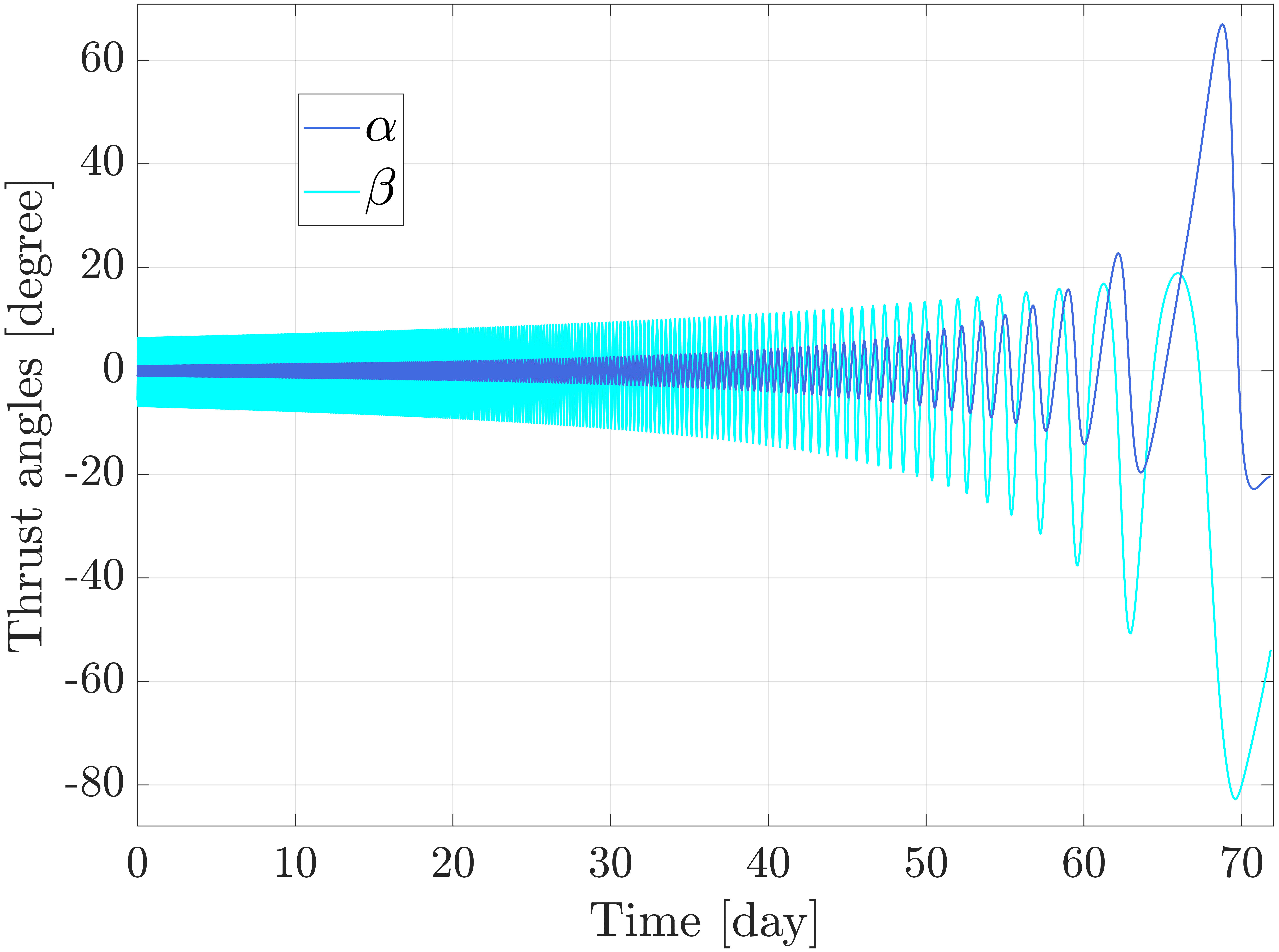}
    \vspace*{-6.9mm}
    \caption{Time history of geocentric thrust angles}
    \label{Fig.alphabetaE}
\end{figure}

\begin{figure}
    \centering
    \includegraphics[width=0.49\textwidth]{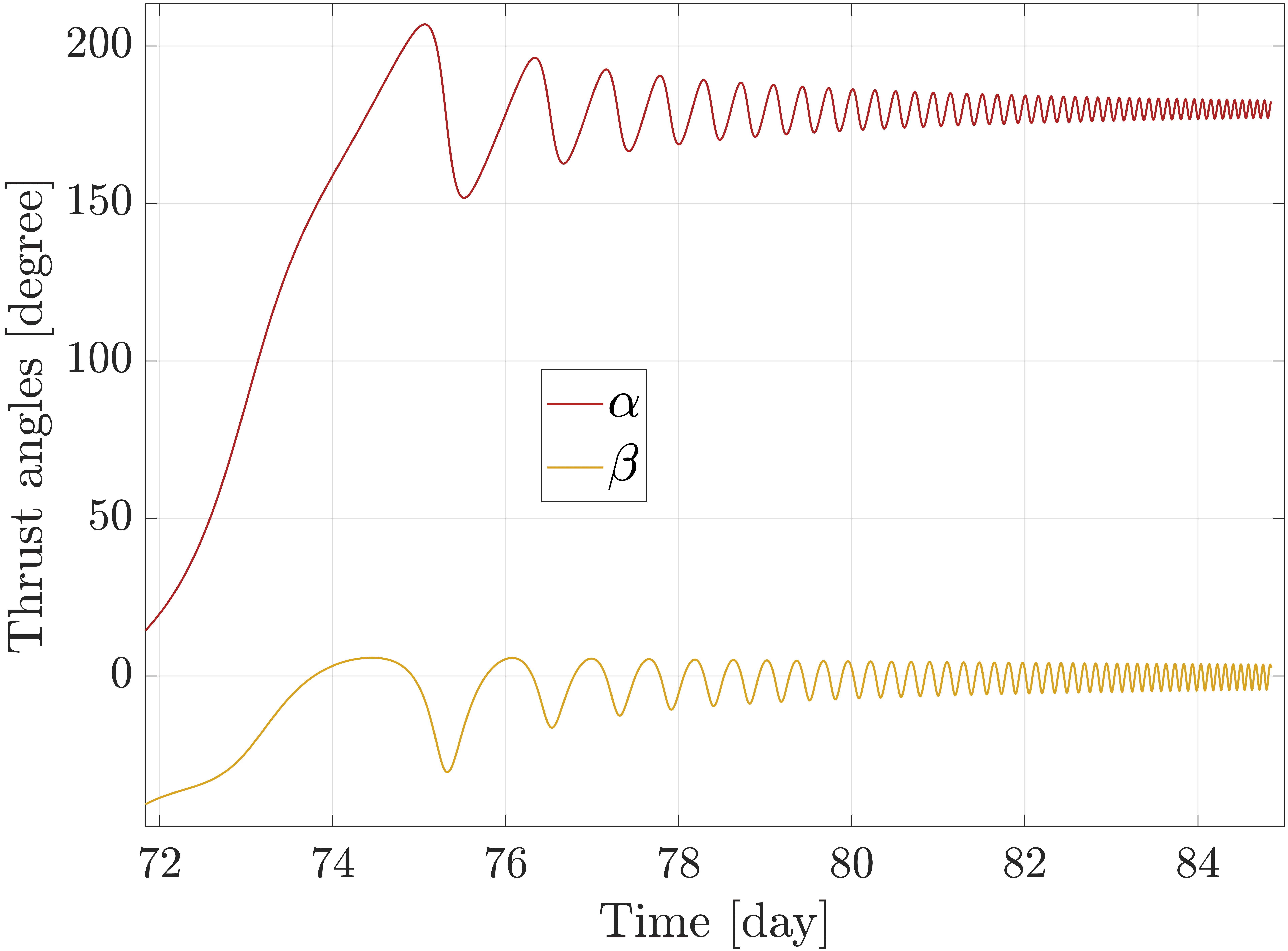}
    \vspace*{-6.9mm}
    \caption{Time history of selenocentric thrust angles}
    \label{Fig.alphabetaM}
\end{figure}

\begin{figure}
    \centering
    \includegraphics[width=0.49\textwidth]{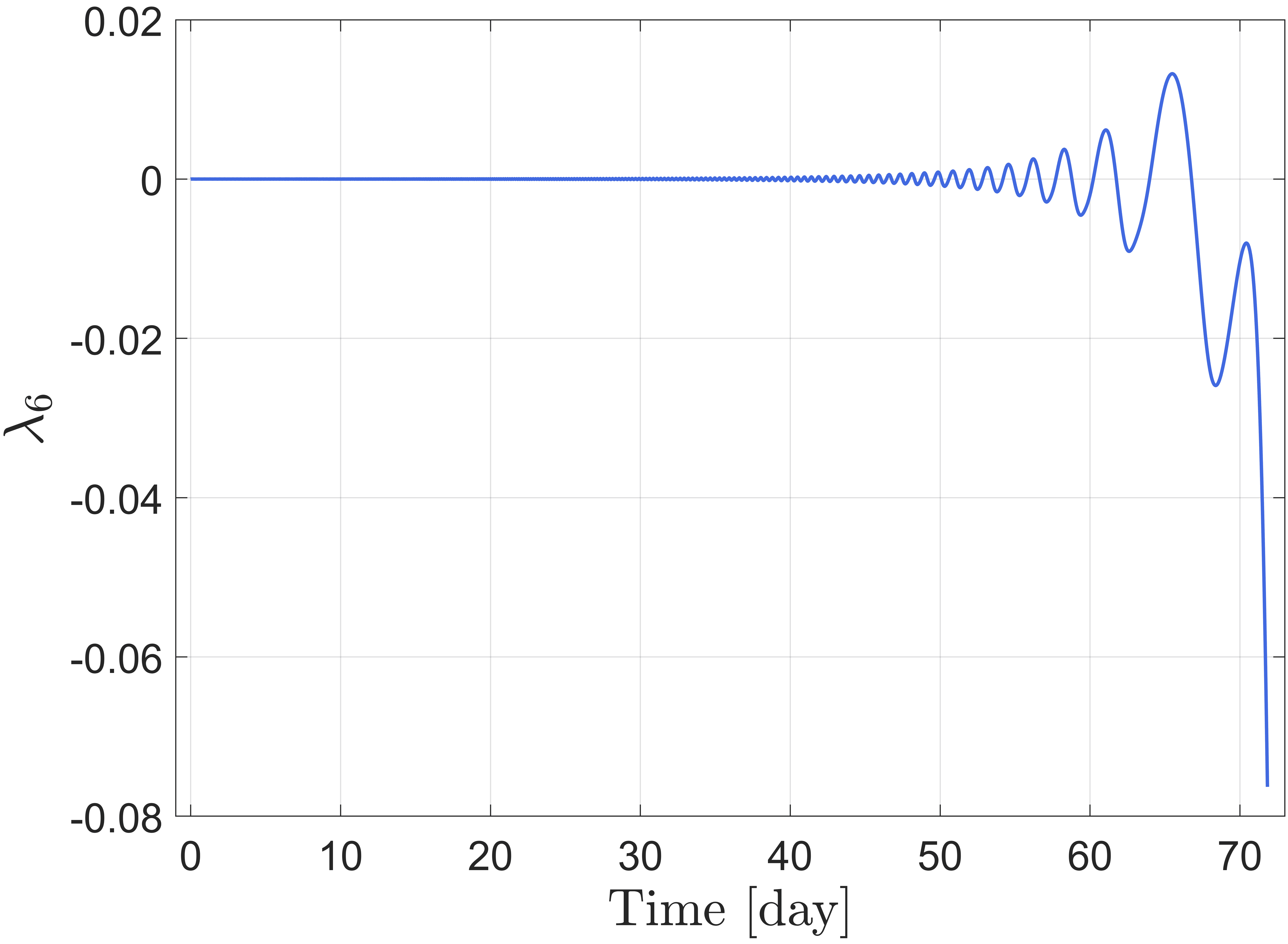}
    \vspace*{-6.9mm}
    \caption{{Time history of $\lambda_6$} in the geocentric leg} 
    \label{Fig.l6GEO}
\end{figure}

\begin{figure}
    \centering
    \includegraphics[width=0.49\textwidth]{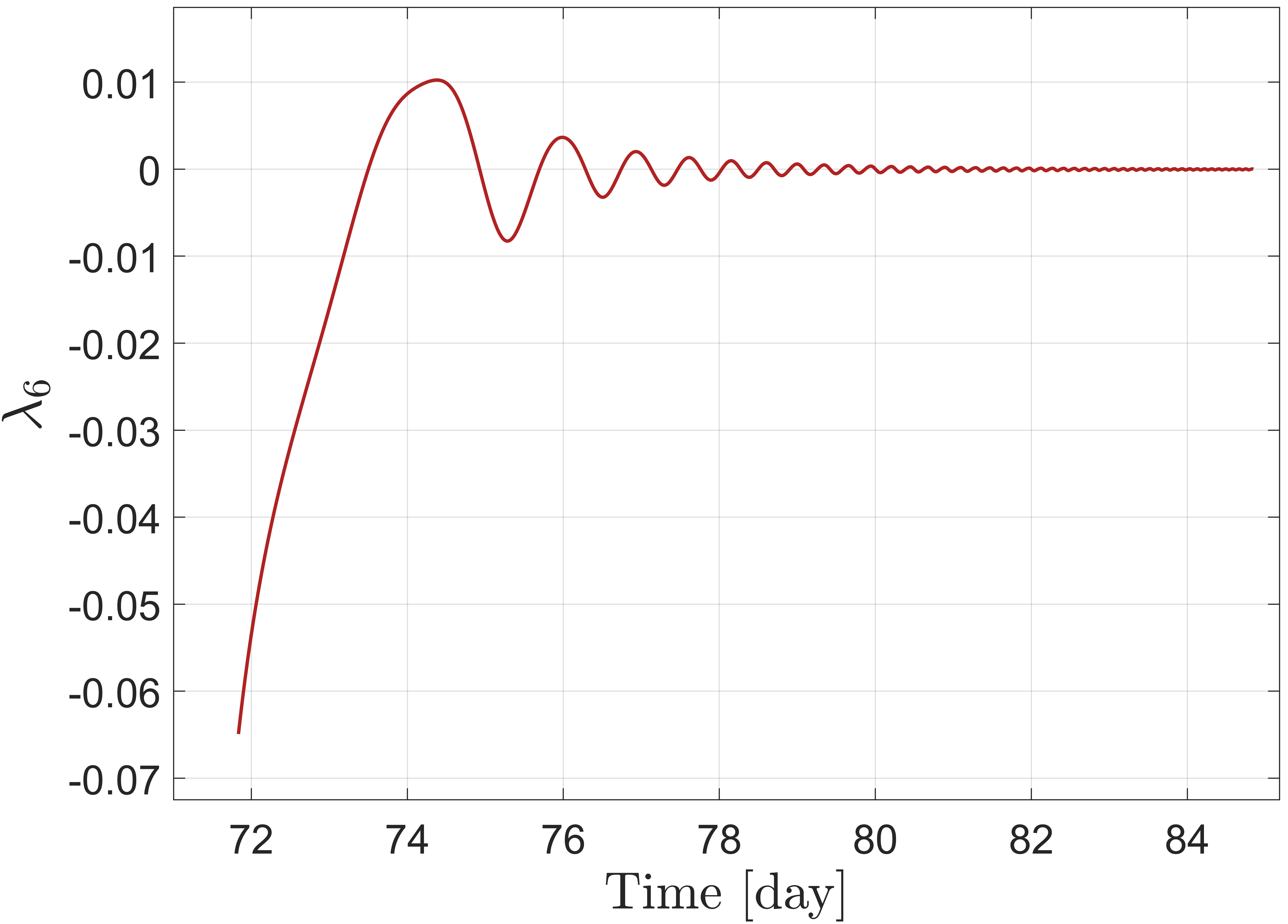}
    \vspace*{-6.9mm}
    \caption{{Time history of $\lambda_6$} in the selenocentric leg} 
    \label{Fig.l6SEL}
\end{figure}

\begin{figure}
    \centering
    \includegraphics[width=0.49\textwidth]{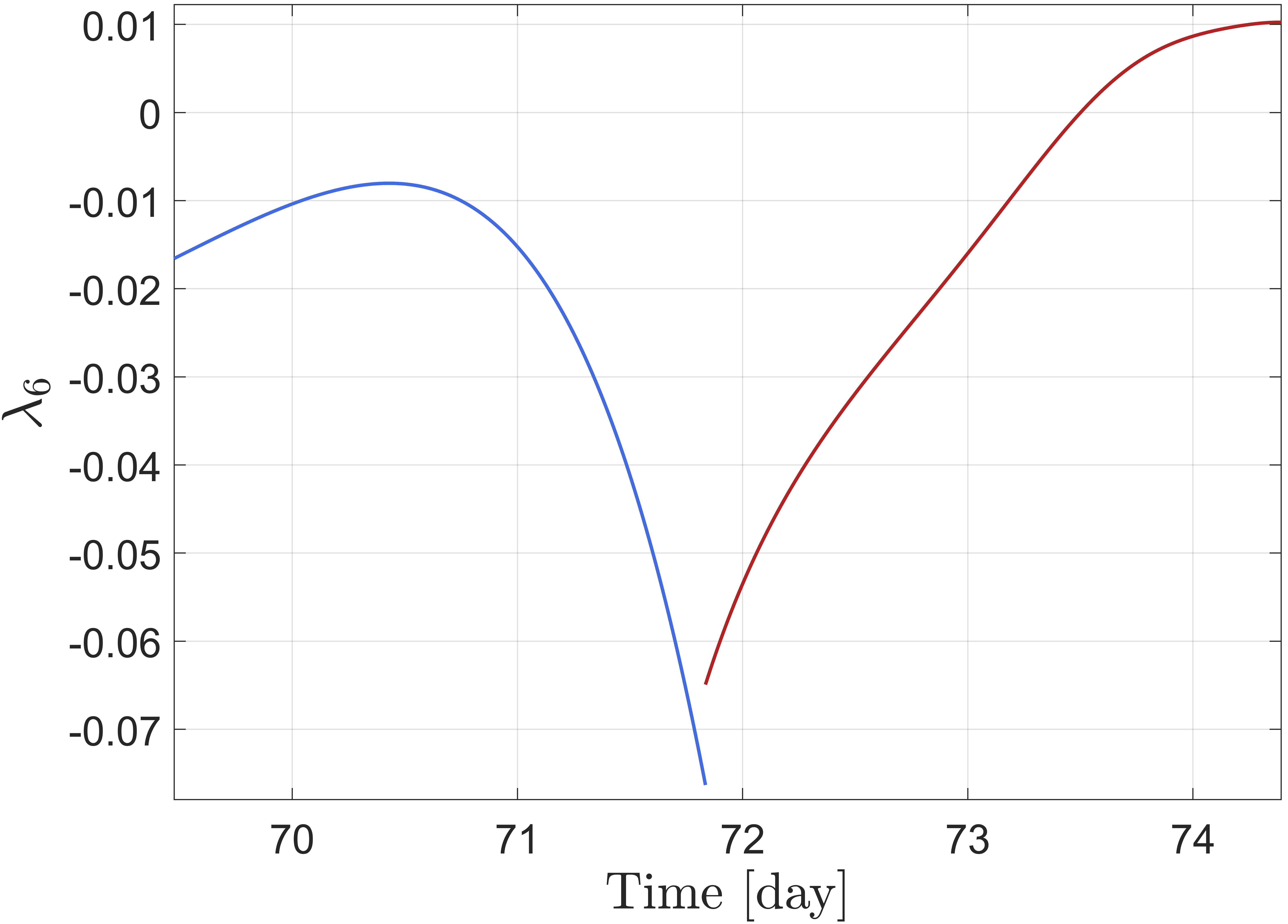}
    \vspace*{-6.9mm}
    \caption{{Zoom on the discontinuity of $\lambda_6$ at the interface}} 
    \label{Fig.l6GEOSEL}
\end{figure}

\begin{figure}
    \centering
    \includegraphics[width=0.49\textwidth]{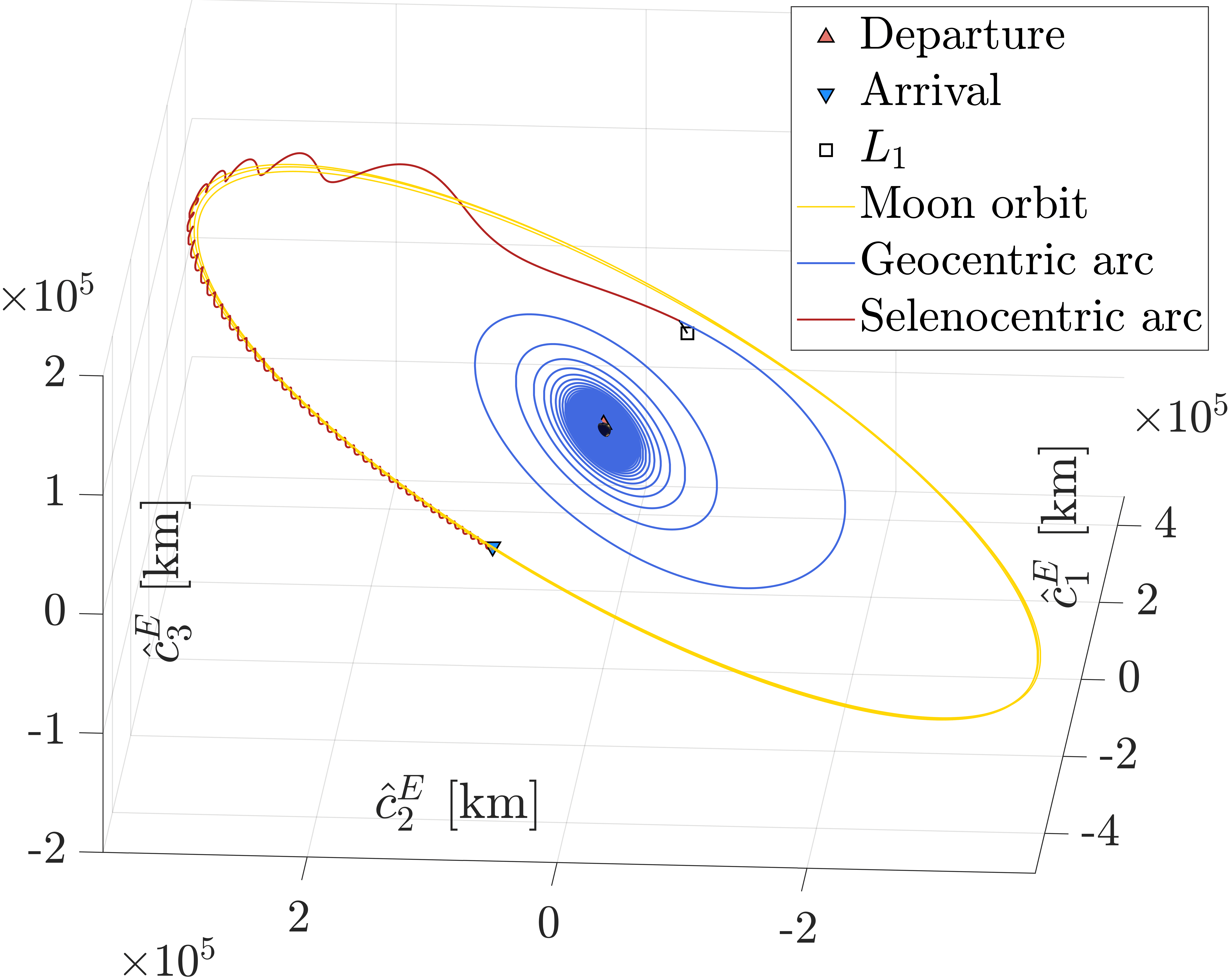}
    \vspace*{-6.9mm}
    \caption{Minimum-time orbit transfer in ECI J2000}
    \label{Fig.pathECI}
\end{figure}

\begin{figure}
    \centering
    \includegraphics[width=0.49\textwidth]{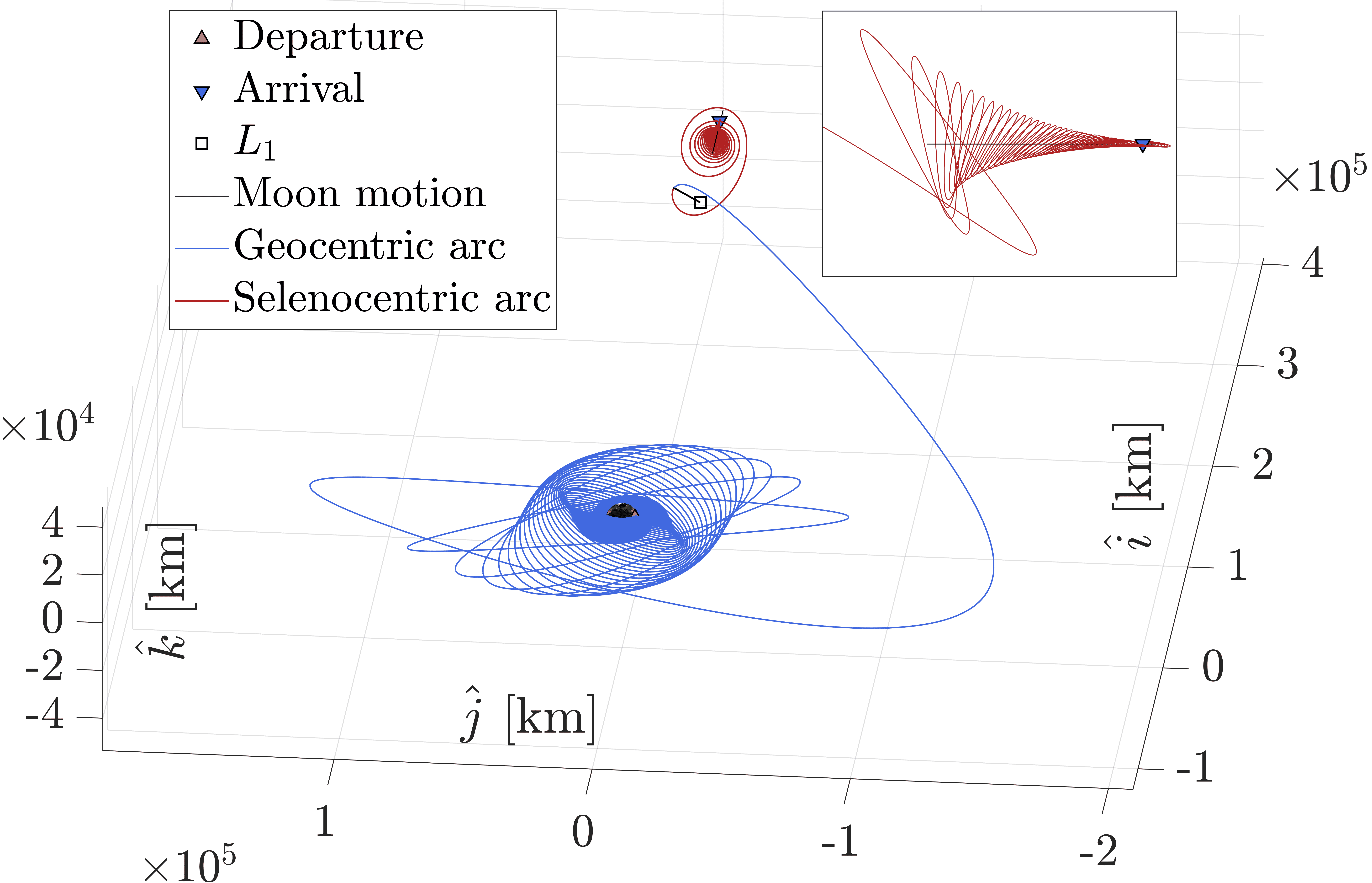}
    \vspace*{-6.9mm}
    \caption{Minimum-time orbit transfer in the Earth-centered synodic reference frame}
    \label{Fig.pathSYN}
\end{figure}

\begin{figure}
    \centering
    \includegraphics[width=0.49\textwidth]{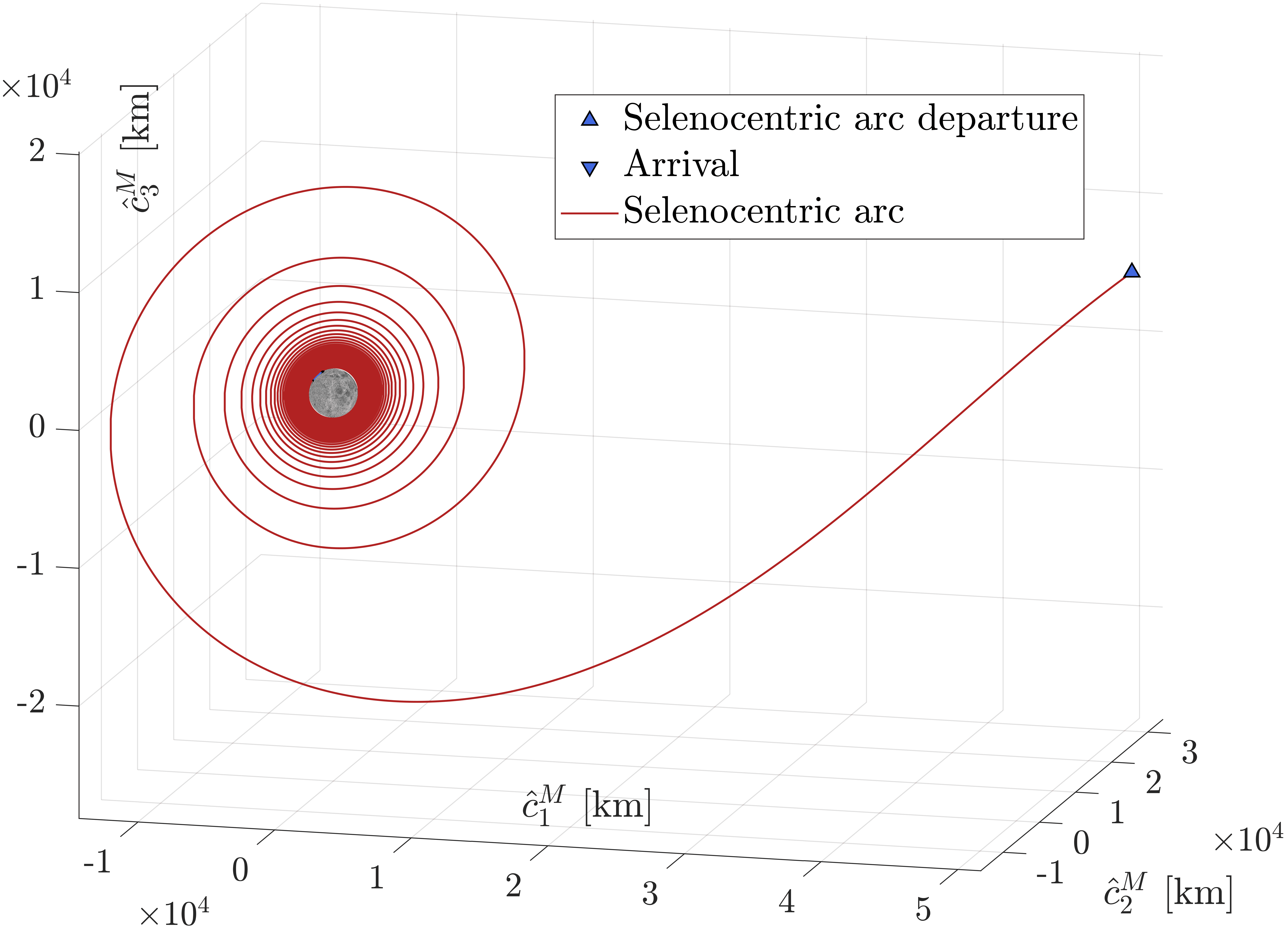}
    \vspace*{-6.9mm}
    \caption{{Selenocentric portion of the minimum-time orbit transfer in MCI}}
    \label{Fig.pathMCI}
\end{figure}

\noindent Similarly, Fig. \ref{Fig.alphabetaE} portrays the geocentric optimal thrust angles associated with $\underline{\underline{LVLH}}_E$, whereas Fig. \ref{Fig.alphabetaM} illustrates the selenocentric optimal thrust angles associated with $\underline{\underline{LVLH}}_M$. To display the discontinuity in the costate variables at the transition between the two legs, the time history of $\lambda_6$ is portrayed in Figures \ref{Fig.l6GEO} and \ref{Fig.l6SEL} for the geocentric and selenocentric leg, respectively. A zoom on the jump in the costate variable is illustrated in Fig. \ref{Fig.l6GEOSEL}. Figures \ref{Fig.pathECI} and \ref{Fig.pathSYN} depict the complete minimum-time transfer path in the ECI J2000 reference frame and in the synodic reference frame centered at the Earth, respectively. Lastly, Fig. \ref{Fig.pathMCI} shows the lunar-captured portion of the trajectory in the MCI reference frame, displaying the inward spiraling motion around the Moon. 

\indent Spiraling motion is ubiquitous when low thrust is employed. This is discernible in the vicinity of both the Earth and the Moon in Fig. \ref{Fig.pathECI}. However, because the transfer trajectory is portrayed in ECI J2000, the selenocentric segment of the minimum-time path appears to be helical due to the orbital motion of the Moon. Also, it is clearly visible that the selenocentric leg takes almost half of the orbit period of the Moon.

\subsection{Specified final RAAN}

\indent If the RAAN of the lunar orbit is specified, Algorithm 2 is employed. The minimum set of parameters is retrieved more easily, without the introduction of any additional parameter,
\begin{equation}\label{MinParSetB}
    \resizebox{.88\hsize}{!}{$\boldsymbol{\Upsilon} = \begin{bmatrix}
        t_{f} & x_{M_{6,f}} & \lambda_{{M_{1,f}}} & \lambda_{{M_{2,f}}} & \lambda_{{M_{3,f}}} & \lambda_{{M_{4,f}}} & \lambda_{{M_{5,f}}}
    \end{bmatrix}^T.$}
\end{equation}
\noindent Each individual corresponds to a specific selection of these unknown parameters. The lower and upper bounds of the search space of each parameter are collected in two vectors, 
\begin{equation}
\resizebox{.88\hsize}{!}{$\begin{split}
    \boldsymbol{LB}_{\Upsilon} &= \begin{bmatrix}
      t_{f}^{(min)} & -\pi & \lambda_{min} & \lambda_{min} & \lambda_{min} & \lambda_{min} & \lambda_{min}
    \end{bmatrix}^T \\ 
    \boldsymbol{UB}_{\Upsilon} &= \begin{bmatrix}
      t_{f}^{(max)} & \;\;\pi & \lambda_{max} & \lambda_{max} & \lambda_{max} & \lambda_{max} & \lambda_{max}
    \end{bmatrix}^T
\end{split}$}
\end{equation}
\noindent where $t_{f}^{(min)}$, $t_{f}^{(max)}$, $\lambda_{min}$, and $\lambda_{max}$ are defined exactly as specified after Eq. (\ref{LBUBFreeRAAN}).

\indent The final state and costate vectors are expressed as
\begingroup
\thinmuskip=0mu
\medmuskip=0mu
\thickmuskip=0mu
\begin{equation}\label{xMf} 
\resizebox{.88\hsize}{!}{$\begin{split}
    \boldsymbol{x}_{M,f} &= \begin{bmatrix}
        p_{M,f} & 0 & 0 & \tan{\frac{i_{M,f}}{2}} c_{\Omega_{M,f}} & \tan{\frac{i_{M,f}}{2}} s_{\Omega_{M,f}} & x_{M_{6,f}}
    \end{bmatrix}^T \\
    \boldsymbol{\lambda}_{{M,f}} &= \begin{bmatrix}
        \lambda_{{M_{1,f}}} & \lambda_{{M_{2,f}}} & \lambda_{{M_{3,f}}} & \lambda_{{M_{4,f}}} & \lambda_{{M_{5,f}}} & 0
    \end{bmatrix}^T.
\end{split}$}
\end{equation}
\endgroup
\noindent It is easy to verify that this representation for the final state and costate vectors satisfies the necessary conditions at the final time (\ref{2family}) and (\ref{lambdabndf2}).

\indent The equality necessary conditions at the initial time (\ref{zeta0}) and (\ref{lambdabnd0}) are collected in the auxiliary vector $\boldsymbol{Q}$, 
\begin{equation}\label{QB}
    \boldsymbol{Q} = \begin{bmatrix}
        x_{E_{1,0}} - p_{E,0} \\ x_{E_{2,0}}^2 + x_{E_{3,0}}^2 - e_{E,0}^2 \\ x_{E_{4,0}}^2 + x_{E_{5,0}}^2 - \tan^2{\cfrac{i_{E,0}}{2}} \\ \lambda_{{E_{2,0}}} \, x_{E_{3,0}} - \lambda_{{E_{3,0}}} \, x_{E_{2,0}} \\ \lambda_{{E_{4,0}}} \, x_{E_{5,0}} - \lambda_{{E_{5,0}}} \, x_{E_{4,0}} \\ \lambda_{{E_{6,0}}}
    \end{bmatrix}.
\end{equation}
\noindent Thus, fitness function $\Gamma_3$ employed by the backward numerical solution method is
\begin{equation}\label{Gamma3B}
    \resizebox{.88\hsize}{!}{$\Gamma_3 \left( \boldsymbol{\Upsilon} \right) = \sqrt{w_1 Q_1^2 + w_2 Q_2^2 + w_3 Q_3^2 + w_4 Q_4^2 + w_5 Q_5^2 + w_6 Q_6^2}$}
\end{equation}
\noindent where $Q_i$ (with $i = 1, \dots, 6$) is the 
$i$-th component of vector $\boldsymbol{Q}$, and the weighting coefficients $w_i$ are all set to 1.

\indent Four distinct selections of $\Omega_{M,f}$ are considered (i.e., \{$0$, $90$, $180$, $270$\} \degree). This time, the numerical solution process employs 100 individuals. The results for each value of the RAAN of the lunar orbit are summarized in Table \ref{TABLE2}. As expected, the time of flight of each of the minimum-time orbit transfers with specified final RAAN is higher than that of the minimum-time path with free final RAAN. In particular, $\Omega_{M,f} = 90$ \degree experiences the greatest time penalty,

\begin{table}[H]
\centering
\caption{Performance and departure RAAN of minimum-time paths for different values of the final lunar RAAN}
\vspace*{-1.5mm}
\begin{tblr}{Q[c,m]|Q[c,m]|Q[c,m]|Q[c,m]|Q[c,m]}
$\Omega_{M,f}$ & $\Omega_{E,0}$ & $\Gamma_3^{(opt)}$ & $t_f^{(opt)}$ & $m_{R,f}^{(opt)}$ \\
$[\degree]$ & $[\degree]$ & $[-]$ & $[\text{day}]$ & $[-]$ \\
\hline
Free & $-22.49$ & $4.02 \cdot 10^{-7}$ & $84.85$ & $0.760$\\ 
$0$ & $5.97$ & $2.71 \cdot 10^{-4}$ & $90.34$ & $0.745$\\
$90$ & $-23.03$ & $2.64 \cdot 10^{-4}$ & $90.63$ & $0.744$\\
$180$ & $-13.37$ & $2.63 \cdot 10^{-4}$ & $85.63$ & $0.758$\\
$270$ & $-7.50$ & $3.63 \cdot 10^{-4}$ & $89.23$ & $0.748$
\end{tblr}
\label{TABLE2}
\end{table}

\begin{figure}[H]
    \centering
    \includegraphics[width=0.49\textwidth]{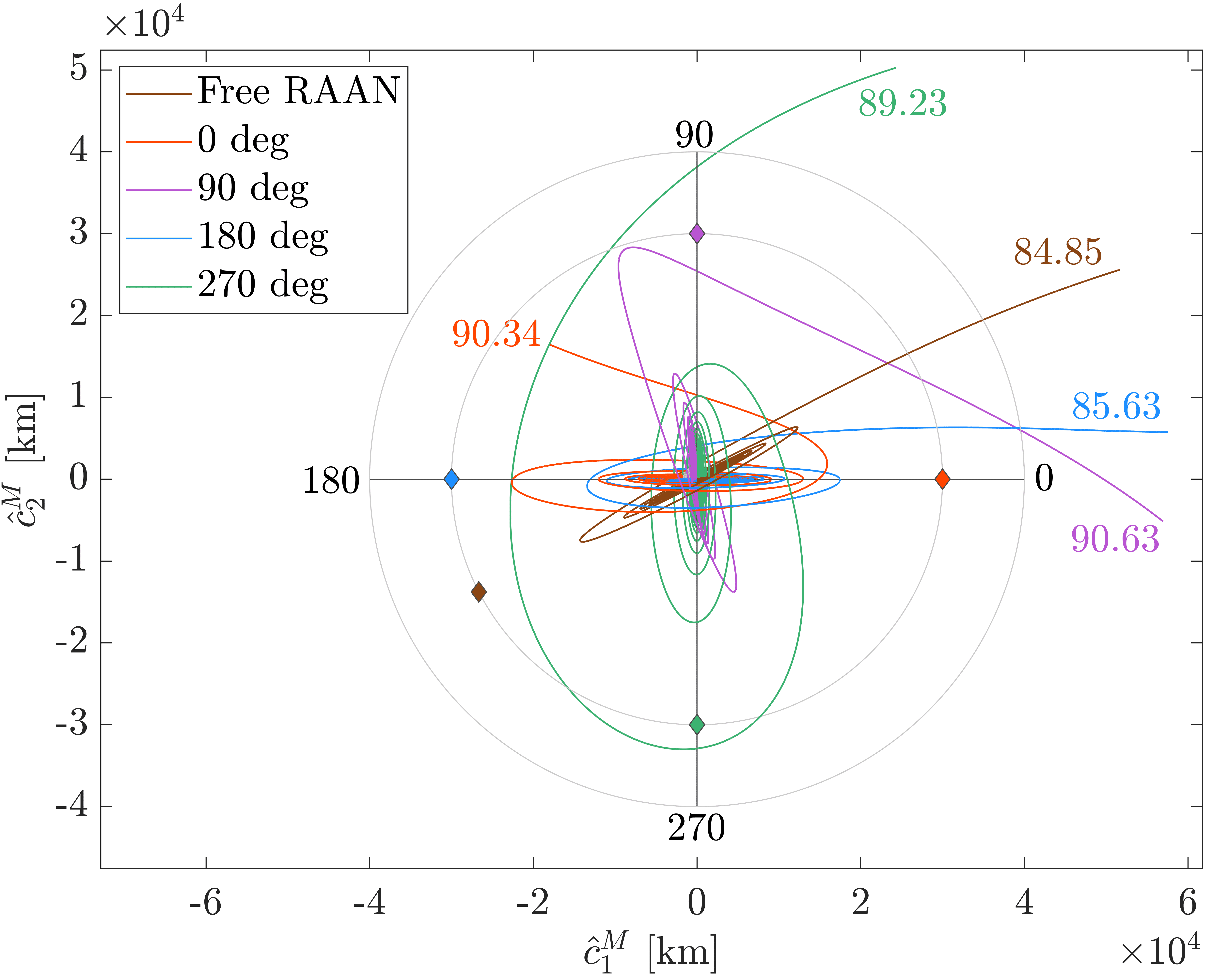}
    \vspace*{-6.9mm}
    \caption{{Planar (Moon equatorial) projection of the } selenocentric legs of the minimum-time paths} 
    \label{Fig.SelLegs}
\end{figure}
\begin{figure}[H]
    \centering
    \includegraphics[width=0.49\textwidth]{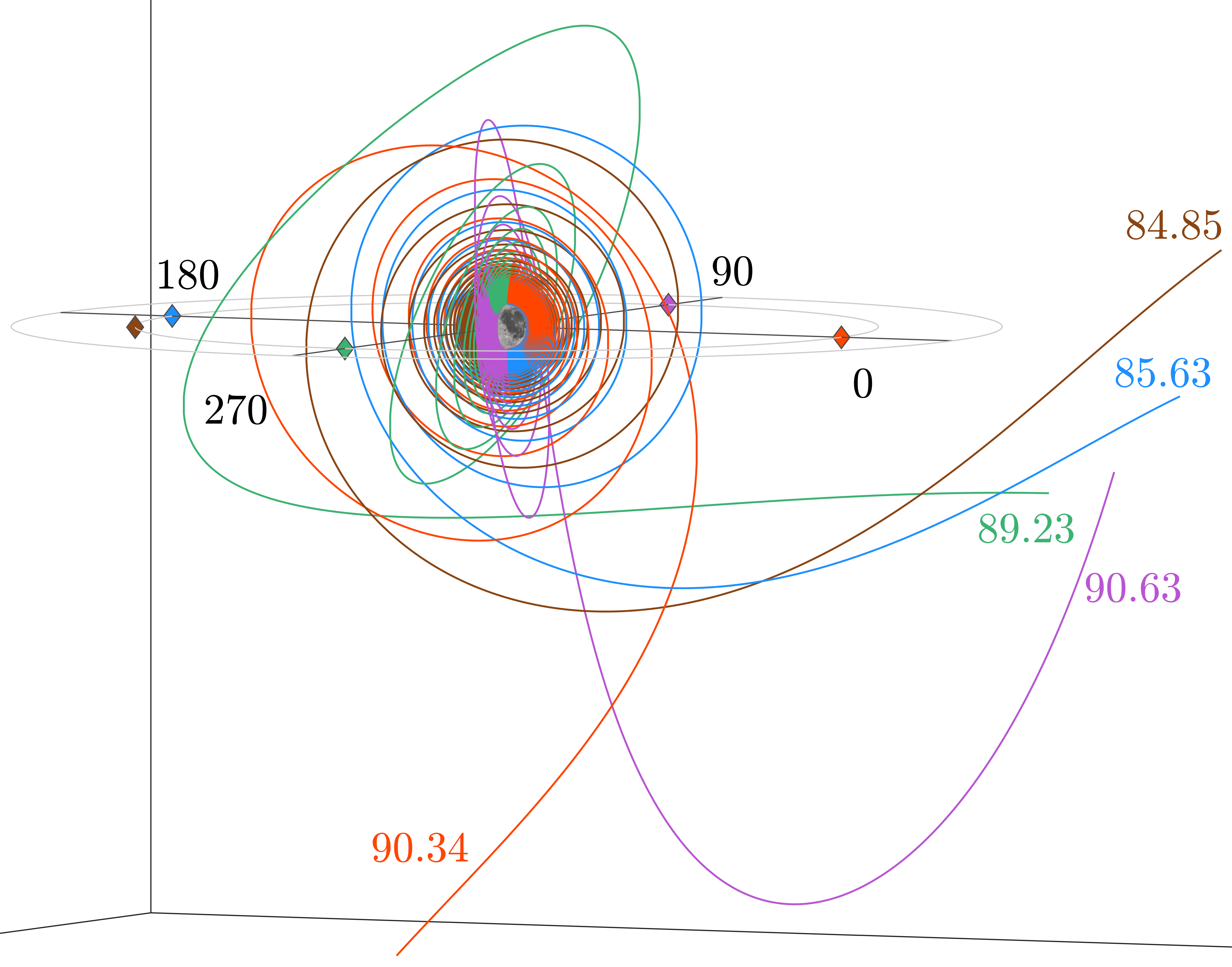}
    \vspace*{-6.9mm}
    \caption{{Three-dimensional view of the selenocentric legs of the minimum-time paths}} 
    \label{Fig.SelLegs3d}
\end{figure}


\noindent followed by $\Omega_{M,f} = 0$ \degree, $\Omega_{M,f} = 270$ \degree, and $\Omega_{M,f} = 180$ \degree. The differences in terms of transfer time among different solutions can be more clearly interpreted by considering that the solution with free final RAAN ultimately settles in the desired lunar orbit with final RAAN of $207.3$ \degree. In fact, when lunar orbits with specified $\Omega_{M,f}$ are targeted, it seems reasonable to expect a longer time of flight as the angular displacement between the desired final RAAN and the optimal (free) RAAN increases. Figure \ref{Fig.SelLegs} portrays the planar projection of the three-dimensional selenocentric legs of the five minimum-time paths in the equatorial plane of the MCI reference frame, and provides a clear visualization of the angular displacement associated with each solution with specified final RAAN. {In addition, a three-dimensional view of the selenocentric legs is presented in Fig. \ref{Fig.SelLegs3d}, for the primary purpose of highlighting the different ways in which each solution approaches the orbital plane of the associated target orbit.} 

\section{CONCLUDING REMARKS}

\indent This study addresses minimum-time low-thrust transfers from low Earth orbit to low lunar orbit, extending the indirect formulation of minimum-time low-thrust orbit transfers about a single attracting body to a framework in which initial and final orbits are placed around two distinct primaries. The underlying optimization problem is very challenging, because during the orbit transfer the spacecraft transitions from orbiting the Earth to orbiting the Moon. This complication is addressed by formulating the problem at hand as a multiple-arc optimization problem, consisting of a geocentric leg followed by a selenocentric leg. The multiple-arc approach requires switching between coexisting representations for the dynamical state of the spacecraft at the transition between consecutive arcs. Correspondingly, a matching relation for the costate, in the form of implicit costate transformation, is derived and enforced at the junction between the two arcs. Modified equinoctial elements are employed to model orbit dynamics, while Cartesian coordinates play the role of convenient intermediate matching variables. All the necessary conditions for optimality are derived, including the multipoint necessary conditions at the junction time between the two arcs. The latter relations are combined, to yield the matching (jump) relation for the costate variables between the two arcs, through a sequence of implicit costate transformations. All the multipoint conditions are proven to admit a closed-form, unique solution and are solvable sequentially. As a result, the parameter set for an indirect algorithm retains the size of the typical set associated with a single-arc optimization problem. Unlike previous studies that use simplified models (such as the circular restricted three-body problem), this research addresses the optimization of Earth-Moon orbit transfers in a high-fidelity dynamical framework, with the use of planetary ephemeris and the inclusion of the simultaneous gravitational action of Sun, Earth, and Moon along the entire transfer path. It is worth remarking that the general four-body model (Earth, Moon, Sun, and spacecraft) captures the essence of the dynamics of the problem, but simplifies the description of the motion of the perturbing bodies. Although one expects to obtain similar results, such framework is suitable in a preliminary mission analysis context, while the methodology proposed in this study can be successfully adopted in a more advanced design stage. With regard to the numerical solution, this work describes and employs the indirect heuristic technique, based on the joint use of all the necessary conditions for optimality, a heuristic algorithm (i.e., differential evolution), and a layered fitness function, aimed at facilitating convergence. Two challenging Earth-Moon low-thrust orbit transfers, (a) with free final right ascension of the ascending node and (b) with specified final right ascension of the ascending node, are selected as illustrative examples. For case (b), backward propagation is shown to be a very convenient option to expedite the convergence of the numerical solution process. For both scenarios (a) and (b), the multiple-arc formulation, in conjunction with the indirect heuristic technique, successfully detects the minimum-time transfers, meeting all the orbit injection constraints and the necessary conditions for optimality, to a great accuracy. 

\section*{{APPENDIX A}}

\indent This appendix is concerned with the derivation of the implicit costate transformation and with the proof of its applicability to the investigated mission scenario. Expanding the left-hand side of Eqs. (\ref{ext1})-(\ref{ext3}) yields 
\begin{align}
    \frac{\partial \Phi}{\partial \boldsymbol{x}_{ini}^{(j + 1)}} &= \boldsymbol{\nu}_j^{T} \frac{\partial \boldsymbol{\chi}_j}{\partial \boldsymbol{x}_{ini}^{(j + 1)}} + \xi_j \frac{\partial \psi_j}{\partial \boldsymbol{x}_{ini}^{(j + 1)}}\label{1}\\
    \frac{\partial \Phi}{\partial \boldsymbol{x}_{fin}^{(j)}} &= \boldsymbol{\nu}_j^{T} \frac{\partial \boldsymbol{\chi}_j}{\partial \boldsymbol{x}_{fin}^{(j)}} + \xi_j \frac{\partial \psi_j}{\partial \boldsymbol{x}_{fin}^{(j)}}\label{2}\\
    \frac{\partial \Phi}{\partial t_{j}} &= \boldsymbol{\nu}_j^{T} \frac{\partial \boldsymbol{\chi}_j}{\partial t_{j}} + \xi_j \frac{\partial \psi_j}{\partial t_{j}}\label{3}
\end{align}
\noindent where $\nicefrac{\partial \boldsymbol{\chi}_j}{\partial \boldsymbol{x}_{fin}^{(j)}}$ and $\nicefrac{\partial \boldsymbol{\chi}_j}{\partial \boldsymbol{x}_{ini}^{(j + 1)}}$ are the $n \times n$ Jacobian matrices of the state matching function with respect to $\boldsymbol{x}_{fin}^{(j)}$ and $\boldsymbol{x}_{ini}^{(j + 1)}$, respectively. These matrices are nonsingular due to the isomorphic (bijective) nature of the state
transformation. The terms $\nicefrac{\partial \psi_j}{\partial \boldsymbol{x}_{fin}^{(j)}}$ and $\nicefrac{\partial \psi_j}{\partial \boldsymbol{x}_{ini}^{(j + 1)}}$ are the ($1 \times n$)-vectors that collect the partial derivatives of the transition function with respect to $\boldsymbol{x}_{fin}^{(j)}$ and $\boldsymbol{x}_{ini}^{(j + 1)}$, respectively, $\nicefrac{\partial \boldsymbol{\chi}_j}{\partial t_{j}}$ is the ($n \times 1$)-vector that contains the partial derivatives of the state matching function with respect to $t_{j}$, and $\nicefrac{\partial \psi_j}{\partial t_{j}}$ is the scalar partial derivative of the transition function with respect to $t_{j}$. 

\indent The transition function determines the switching time from the first to the subsequent arc. For the problem at hand, the transition condition is assumed to be verified when the distance between spacecraft and Moon drops below a certain arbitrary threshold. The physical interpretation of this condition corresponds to the spacecraft entering a SOI of given radius centered at the Moon. The general expression for the transition function is such that the function is greater than zero if the spacecraft is outside the Moon SOI, equal to zero if the spacecraft lies exactly at the contour of the SOI and less than zero if the spacecraft is inside the SOI, 
\begin{equation}\label{psigeneral}
    \resizebox{.88\hsize}{!}{$\psi_1 = d^2 \left( spacecraft, \Moon \right) - \rho_M^2 \begin{cases}
        < 0 \: &\text{inside SOI}\\
        = 0 \: &\text{at SOI bound}\\
        > 0 \: &\text{outside SOI}
    \end{cases}$}
\end{equation}
\noindent where $d \left( spacecraft, \Moon \right)$ indicates the distance between the spacecraft and the Moon, whereas $\rho_{M}$ is the radius of the Moon SOI.

\indent The transition function only concerns the position of the spacecraft. This is the case for many multiple-arc space trajectory optimization problems, for which the condition for transitioning from one arc to the next is only a function of the spacecraft {position \cite{pontani2022optimal}}. Anyway, considering both the position vector from the Earth to the spacecraft and that from the Moon to the spacecraft, two possible formulations exist for $\psi_1$:
\begin{itemize}
    \item transition function depending on the position vector of the spacecraft with respect to the Earth, $\boldsymbol{r}_{E}$, and on the switching time $t_1$ (because the position vector from the Earth to the Moon is provided by the ephemeris model), 
    \begin{equation}\label{psiE}
        \psi_{1_E} \left( \boldsymbol{r}_{E}, t_1 \right)  = \left| \boldsymbol{r}_{E} - \boldsymbol{r}_{\Moon_E} \left( t_1 \right) \right|^2 - \rho_{M}^2
    \end{equation}
    \item transition function depending only on the position vector of the spacecraft with respect to the Moon, $\boldsymbol{r}_{M}$,
    \begin{equation}\label{psiM}
        \psi_{1_M} \left( \boldsymbol{r}_{M} \right) = \left| \boldsymbol{r}_{M} \right|^2 - \rho_{M}^2 = r_M^2 - \rho_{M}^2.
    \end{equation}
\end{itemize}
\noindent It is apparent that the second formulation is convenient, but, for the sake of completeness, both expressions are presented. 
\indent It is rather reasonable to argue that the switching function can be set up arbitrarily, with a randomly chosen distance from the Moon as the triggering threshold for the transition from the first to the second arc. This can be proven via \textit{reductio ad absurdum} by assuming that an optimal distance from the Moon for the transition to occur does exist. Thus, this distance can be regarded as a component of the time-independent parameter vector $\boldsymbol{p}$ introduced in the formulation of the multiple-arc optimal control problem. Considering either one of the two formulations for $\psi_1$ and the quantity $\rho_M$ as a time-independent optimization parameter, the necessary condition on the parameter vector presented in Eq. (\ref{pnecconds}) yields 
\begin{equation}\label{DIM} 
    \frac{\partial \Phi}{\partial \rho_M} = \frac{\partial \psi_1}{\partial \rho_M} = -2 \, \xi_1 \, \rho_M = 0.
\end{equation} 
\noindent Because $\rho_M = 0$ is clearly unfeasible (zero distance from the center of the Moon), Eq. (\ref{DIM}) entails that the time-independent adjoint conjugate to the switching function is identically zero (i.e., $\xi_1 = 0$), or equivalently that the transition function $\psi_1$ can be removed from the problem formulation. This demonstrates unequivocally that the distance from the Moon at which the transition occurs is arbitrary, and ultimately represents an additional degree of freedom for the design of the solution strategy. 

\indent Using Eqs. (\ref{ext1}) and (\ref{ext2}), together with $\xi_1 = 0$, Eqs. (\ref{1}) and (\ref{2}) can be rewritten as 
    \begin{align}
    -\boldsymbol{\lambda}_{ini}^{(j + 1)^T} &= \boldsymbol{\nu}_j^{T} \frac{\partial \boldsymbol{\chi}_j}{\partial \boldsymbol{x}_{ini}^{(j + 1)}}\label{11}\\
    \boldsymbol{\lambda}_{fin}^{(j)^T} &= \boldsymbol{\nu}_j^{T} \frac{\partial \boldsymbol{\chi}_j}{\partial \boldsymbol{x}_{fin}^{(j)}}.\label{21}
\end{align}
\noindent Lastly, transposition and combination of Eqs. (\ref{11}) and (\ref{21}) yields the implicit costate transformation presented in Eq. (\ref{CostateImplicit}).

\section*{{APPENDIX B}}


\indent This appendix presents the analytical approximation that was adopted to obtain a reliable estimate of the transfer time for the low-thrust orbit transfer of interest. The following analytical approach, developed for two-dimensional, circle-to-circle low-thrust orbit transfers (with a single-attracting-body) can be used. The analytical solution relies on two further approximations: 
\begin{itemize}
    \item the thrust is directed along the velocity vector at all times ($\underrightarrow{\boldsymbol{T}} \parallel \underrightarrow{\boldsymbol{v}}$);
    \item the small nonzero eccentricity values reached during the transfer due to low thrust are neglected, and $e$ is set  to $0$ for the entire transfer duration.
\end{itemize}
\noindent Moreover, hence forward, the final orbit radius $r_f$ is assumed to be greater than the initial orbit radius $r_0$. The rate of the specific energy $\mathcal{E}$ is
\begin{equation}
    \dot{\mathcal{E}} = \frac{\underrightarrow{\boldsymbol T}}{m} \cdot \underrightarrow{\boldsymbol v} = a_T \, v
\end{equation}
\noindent where $a_T = \nicefrac{T}{m}$. The following relation holds for the specific orbital energy:
\begin{equation}
    \mathcal{E} = -\frac{\mu}{2 \, a}
\end{equation}
\noindent where $\mu$ denotes the gravitational parameter of the attracting body, whereas $a$ is the osculating semimajor axis of the transfer path. Differentiating the previous equation with respect to time one gets 
\begin{equation}\label{EpsDot}
    \dot{\mathcal{E}} = \frac{\mu}{2 \, a^2} \, \dot{a} = a_T \, v.
\end{equation}
\noindent Because the eccentricity is set to zero, the transfer trajectory is instantaneously approximated as circular. Consequently, $a$ corresponds to the radius of the circular orbit and the velocity has the typical expression holding for circular orbits,
\begin{equation}\label{AppVel}
    v \simeq \sqrt{\frac{\mu}{a}}.
\end{equation}
\noindent By substituting Eq. (\ref{AppVel}) in the rightmost term presented in Eq. (\ref{EpsDot}) one obtains
\begin{equation}
    \dot{a} = \frac{2 \, a_T}{\sqrt{\mu}} \, a^{\frac{3}{2}}.
\end{equation}
\noindent Separation of variables yields
\begin{equation}\label{Int1}
    \int_{r_0}^{r_f} \frac{1}{2 \, a^{\frac{3}{2}}} {\rm d} a = \int_{t_0^a}^{t_f^a} \frac{a_T}{\sqrt{\mu}}{\rm d} t.
\end{equation}
\indent Using the notable result $u_T = u_T^{(max)}$ in (\ref{controlxoptj}) and the analytical expressions for the the mass ratio (\ref{x7anal}), the following analytical expression for $a_T$ can be obtained:
\begin{equation}\label{aTanal}
    a_T \left( t \right) = \frac{c \, u_T^{(max)}}{c - u_T^{(max)} \, \left( t - t_{0} \right)}.
\end{equation}
\noindent Then, it is possible to integrate separately both sides of Eq. (\ref{Int1}),
\begin{equation}\label{Int2}
        \frac{1}{\sqrt{r_0}} - \frac{1}{\sqrt{r_f}} = \frac{c}{\sqrt{\mu}} \ln{ \frac{c - u_T^{(max)} \left( t_0^a - t_{0} \right)}{c - u_T^{(max)} \left( t_f^a - t_{0} \right)}}
\end{equation}
\noindent where $t_0^a$ and $t_f^a$ indicate the initial and final time of the analytical approximation of the orbit transfer, respectively. Developing further the previous equation one gets
\begin{equation}\label{Int3}
    \resizebox{.88\hsize}{!}{$t_f^a = t_0^a + \frac{1}{u_T^{(max)}} \left\{ c - \left[ c - u_T^{(max)} \left( t_0^a - t_{0} \right) \right] \rm{e}^{-\cfrac{\sqrt{\mu}}{c} \left( \cfrac{1}{\sqrt{r_0}} - \cfrac{1}{\sqrt{r_f}} \right)} \right\}.$}
\end{equation} 
\indent Equation (\ref{Int3}) can be employed to obtain a first-order estimate of the time of flight. For the sake of the application of this approximate solution, it is necessary to divide the trajectory in two arcs, with the Earth considered as the single attracting body in the first arc and the Moon considered as the single attracting body in the second arc. For the geocentric arc, the initial circular orbit radius is set to $r_{E,0} = p_{E,{0}}$, while the final circular orbit radius is set to $r_{E,f} = \rho_{E_{L_1}}$, where $\rho_{E_{L_1}}$ is the distance of the interior libration point $L_1$ from the Earth center of mass. The initial time $t_{E,0}^a$ is set to $t_{0}$ and application of Eq. (\ref{Int3}) yields $t_{E,f}^a$. For the selenocentric arc, in order to coherently apply the analytical approximation, it is necessary to consider the reversed trajectory (i.e., starting from the arrival orbit). The initial circular orbit radius is set to $r_{M,0} = p_{M,{f}}$, while the final circular orbit radius is set to $r_{M,f} = \rho_{M_{L_1}}$, where $\rho_{M_{L_1}}$ is the distance of the interior libration point $L_1$ from the Moon center of mass. The initial time $t_{M,0}^a$ is set to $t_{E,f}^a$, and application of Eq. (\ref{Int3}) yields $t_{M,f}^a$. The overall transfer time estimate is obtained by summing up the two independent contributions, 
\begin{equation}\label{tfinestimate}
    \begin{split}
        \Delta \hat{t} &= \hat{t}_{f} = t_{E_f}^a - t_{E_0}^a + t_{M_f}^a - t_{M_0}^a = t_{M_f}^a - t_{E_0}^a =\\
        &= 81 \; \text{d}, \; 18 \; \text{h}, \; 16 \; \text{m}, \; 44 \; \text{s}. 
    \end{split}
\end{equation}
\indent The presented analytical method only considers a single attracting body in each arc (neglecting the third-body perturbation), and it is rather reasonable to expect that the transfer time is underestimated by this approach. The reason for this lies in the fact that the analytical approximation neglects the three-dimensionality of the orbit transfer, because the orbital plane changes required for the transfer orbit to reach the desired destination are not taken into account. Thus, the optimal transfer time $t_{f}^{(opt)}$ can be conjectured to be greater than $\hat{t}_{f}$.

\section*{ACKNOWLEDGMENTS}

\indent A. Beolchi, C. Pozzi, and E. Fantino acknowledge financial support from projects 8434000368 (6U Cubesat mission funded by Khalifa University of Science and Technology and Al Yah Satellite Communications Company - Yahsat) and CIRA-2021-65/8474000413 (Khalifa University of Science and Technology’s internal grant).

\indent The authors wish to express their gratitude to Francesco Corallo, for his insightful suggestions, with reference to the numerical solution methodology.

\bibliographystyle{ieeetr}
\bibliography{main.bib}

\end{document}